\documentclass[12pt]{article}
\usepackage{graphicx}
\usepackage{amsmath} 
\usepackage{amssymb}
\usepackage{epsfig}
\usepackage{setspace}
\usepackage{float}
	\usepackage [english]{babel}
	\usepackage [autostyle, english = american]{csquotes}
	\MakeOuterQuote{"}
\newcommand{\degree}{\ensuremath{^\circ}}
\RequirePackage{lineno}
\begin{document}
\title{\hfill \small Draft version, submitted to NIM  November 8, 2018.  
\\
{\bigskip}
\Large \bf  Radiation Tests of Hamamatsu Multi-Pixel Photon Counters}
 \maketitle
\begin{center}
\author{G.~Blazey, J.~Colston, A.~Dyshkant, K.~Francis, J.~Kalnins,  S.~A.~Uzunyan, V.~Zutshi\\ ~{\it Department of Physics, Northern Illinois University, DeKalb, IL 60115, USA};\\
      S.~Hansen, P.~Rubinov,\\ 
      {~\it Fermi National Accelerator Laboratory, Batavia, IL 60510, USA; }\\
      E.~C.~Dukes, Y.~Oksuzian,\\{~\it University of Virginia, Charlottesville, VA 22904, USA}\\
      M.~Pankuch\\{~\it  Northwestern Medicine Proton Center, Warrenville, IL 60555, USA}  }
\end{center}
\section{Abstract\label{abstract}}
Results of radiation tests of Hamamatsu 2.0$\times$2.0~mm$^2$ through-silicon-via (S13360-2050VE) 
multi-pixel photon counters, or MPPCs, are presented~\cite{hamamatsu}.
Distinct sets of eight MPPCs were exposed to four different 1~MeV neutron equivalent doses of 200 MeV protons. 
Measurements of the breakdown voltage, gain and noise rates at 
different bias overvoltages, photoelectron thresholds, and LED illumination levels were taken before and after irradiation.  No significant deterioration in performance was observed for breakdown voltage, gain, and response. Noise rates increased significantly with irradiation.
These studies were undertaken in the context of MPPC requirements for the Cosmic Ray Veto detector of the Mu2e experiment at the Fermi National Accelerator Laboratory.
\section{Introduction\label{intro}}
\begin{figure}[ht]
\centering
\includegraphics[height=50mm]{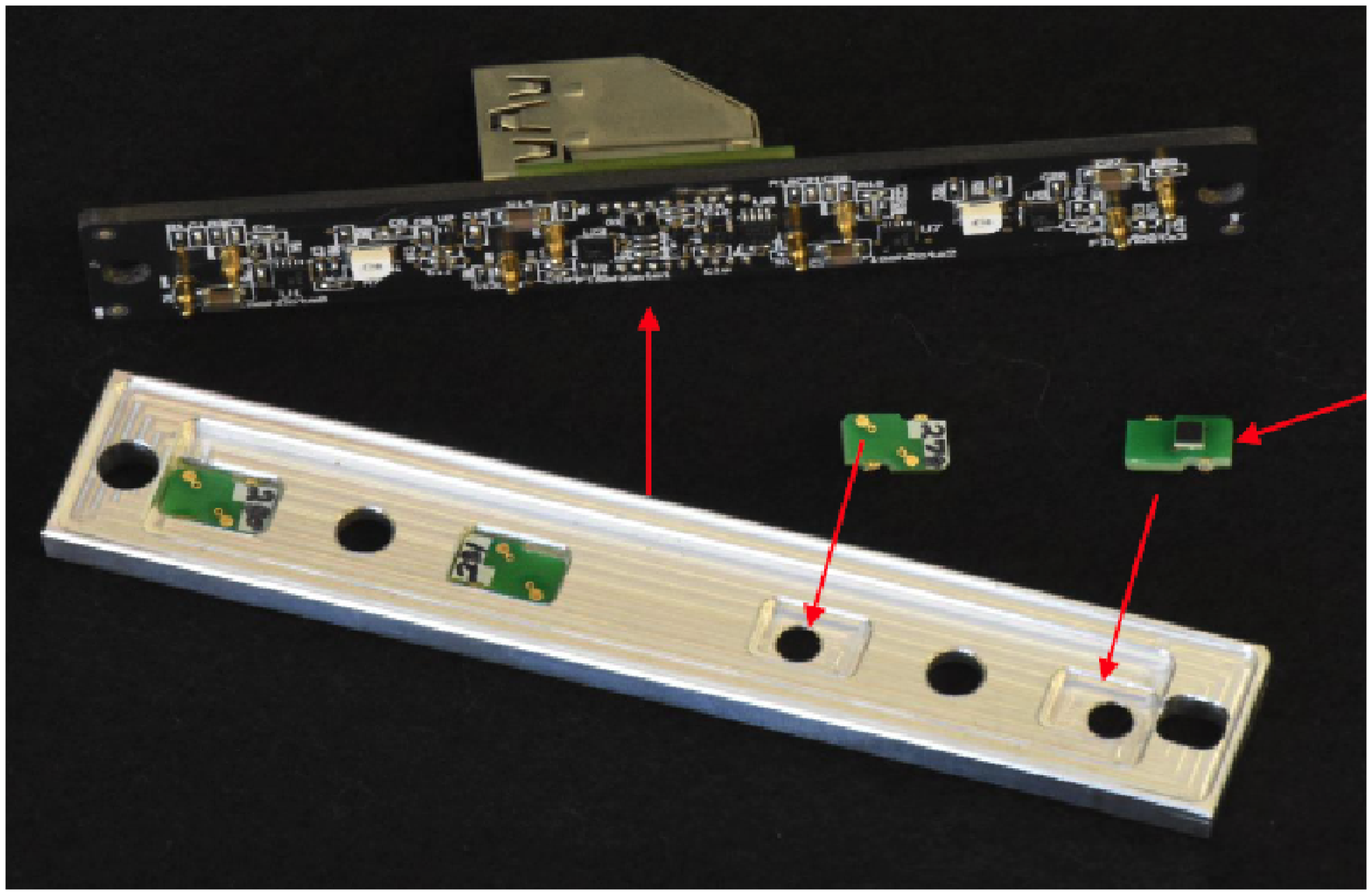}
\includegraphics[height=50mm]{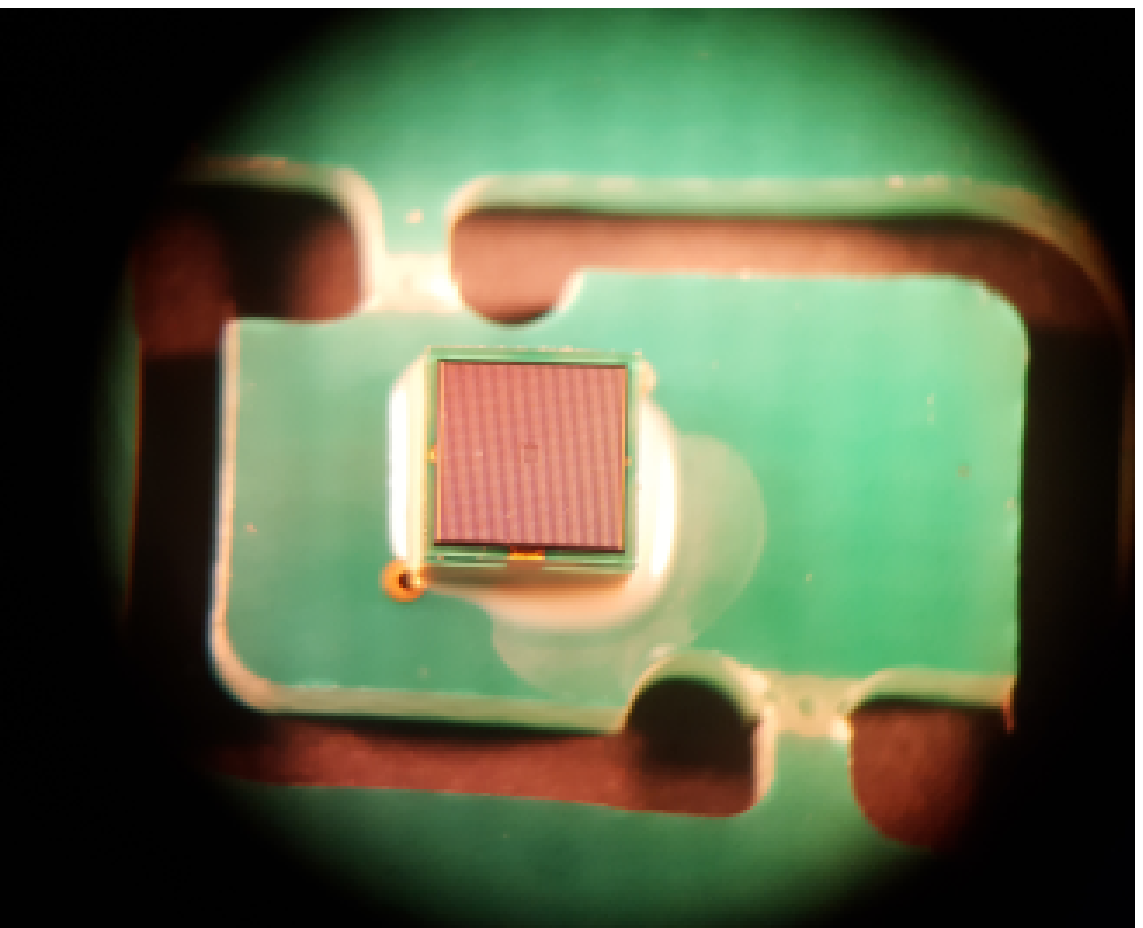}
\caption{\label{fig:sipm_assembly} The MPPC mounting block (left), which accommodates four MPPCs mounted on carrier boards; a closeup of one 2.0$\times$2.0~mm$^2$ S13360-2050VE Hamamatsu MPPC (right) within the carrier board panel.}
\end{figure}

\begin{figure}[ht]
\centering
\includegraphics[scale=0.5]{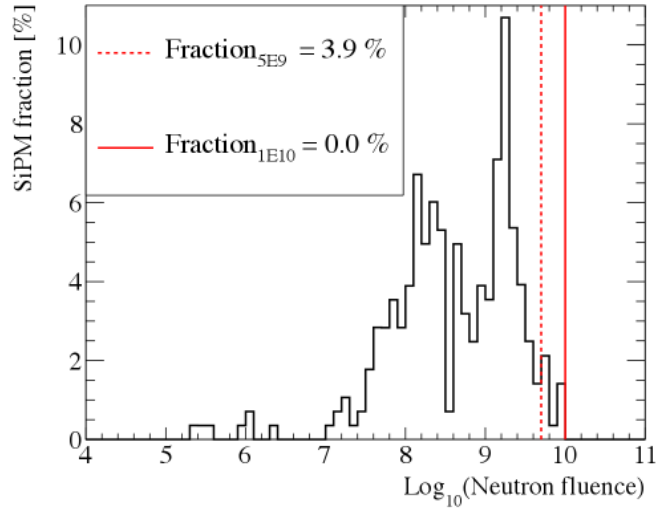}
\caption{\label{fig:neutron_damage} Fraction of CRV MPPCs versus expected radiation fluences over lifetime of Mu2e experiment. Distribution of peaks is a result of the layout of the CRV. }
\end{figure}

The cosmic ray veto (CRV) system of the Mu2e experiment at the Fermi National Accelerator Laboratory (Fermilab)~\cite{Mu2eTDR} is designed to identify incoming cosmic rays with an efficiency of 99.99\% in order to suppress signals from cosmic ray interactions that mimic the muon-to-electron conversion signal. Cosmic ray detection is provided by four 
layers of scintillation counters with embedded wave length shifting (WLS) fibers connected to multi-pixel photon counters (MPPCs) on a  mounting block as shown in Fig.~\ref{fig:sipm_assembly}.
To meet the CRV detection efficiency requirement, summed signals from photodetectors at each end of
a scintillation counter should provide a photoelectron (PE) yield of at least 25~PE/cm  
from a minimum ionizing particle traversing at normal incidence one meter from the counter end. 
The entire system will require 19,840~MPPCs, which are expected to accumulate a maximum radiation fluence of approximately 1$\times$10$^{10}$~neutrons/cm$^2$ from 1 MeV equivalent neutrons over the lifetime of the Mu2e experiment (see Fig.~\ref{fig:neutron_damage}). Note that only a few percent of the MPPCs see fluence above 5$\times$10$^{9}$~ neutrons/cm$^2$. 

%
\begin{table}[h!]
\centering
\caption{Hamamatsu specifications for the 2.0$\times$2.0~mm$^2$ (S13360-2050VE) Through-silicon-via surface mount type MPPCs \cite{hamamatsu}.}
 \begin{tabular}{l p{4cm} l} 
 \hline \hline
 Number of Pixels \hspace{15mm}  & 1584 \\
 Pixel Pitch       & 50 $\mu$m \\
 Response Range  & 320 to 900 nm \\
 Peak Sensitivity  & 450 nm \\
 PDE               & 40\% \\
 Gain              & 1.7$\times$10$^6$\\
 Terminal Capacitance & 140 pF \\
 Dark count rate at 0.5 PE     & 300 KHz \\
 V$_{breakdown}$ & 53V $\pm$ 5V \\
 Bias Voltage   & V$_{breakdown}$+3V \\
 Reference Temperature & 25 $^{\circ}$C \\

 \hline
 \hline
 \end{tabular}
 
 \label{tab:mppc_specs}
\end{table}

We present results of radiation  tests of  Hamamatsu 2.0$\times$2.0~mm$^2$ MPPCs, which are  intended for the instrumentation of the CRV detector. Table \ref{tab:mppc_specs} provides  detailed specifications for the devices. 

The mandated functionality of the CRV over the lifetime of the experiment requires that, 
with the expected accumulated dose, the MPPC a) breakdown voltages do not drift, b) responses do not change, and c) noise rates remain within the data acquisition system 
bandwidth with the application of a threshold that maintains rejection. Additionally, resolution of photoelectron spectra for as great a dose as possible is desirable  for ease of  {\it in situ} calibration and subsequent maintenance of readout thresholds. The requirements are listed in Table \ref{tab:mu2e_specs}.
\begin{table}[h!]
\centering
 \caption{The Mu2e-CRV requirements for the MPPCs.}
 \begin{tabular}{l p{4cm} l} 
 \hline\hline
 Breakdown Voltage Drift \hspace{5mm}  & $<\pm250$~mV \\
 Response Stability       & $<\pm5\%$ \\
 DAQ Bandwidth  & 1 Mhz \\
 DCR Limit     & $<$300 KHz \\	 
 DAQ Threshold  & $<$6 PE \\
 \hline
 \hline
 \end{tabular}

 \label{tab:mu2e_specs}
\end{table}
\section{\label{1p3tests}MPPC Tests}
\begin{figure}[ht]
\centering
\includegraphics[scale=0.6]{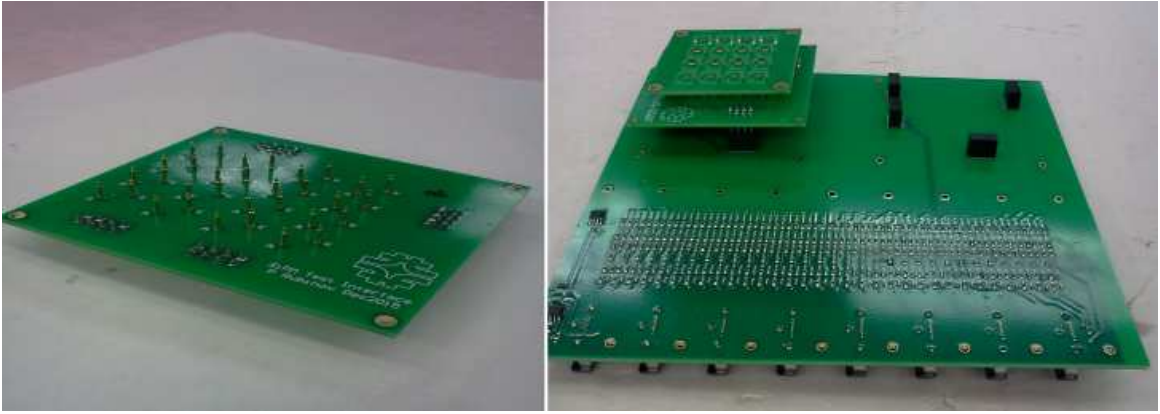}
\caption{\label{fig:sipm_CRboard} The 16-MPPC pogo-board with spring loaded contacts (left) and the MPPC carrier board panel in contact with the pogo-board, mounted on the interface board (right).}
\end{figure}

 To carry out the performance studies, the MPPCs are mounted on small printed circuit boards called carrier board panels (see Fig.~\ref{fig:sipm_assembly} right). These carrier board panels connect to the readout electronics through a pogo-board, a passive board that has an array of spring loaded contacts, or pogo-pins. The pogo-board is then mounted on an interface board which communicates with the Mu2e front-end electronics board (FEB) outside a light-tight box through HDMI cables \cite{FEB}.
 Fig.~\ref{fig:sipm_CRboard}  shows  a 16-MPPC carrier board panel in contact with the pogo-board which in turn is mounted on the interface board.
 MPPCs were placed in a light-tight box to take measurements with and without LED (type LED5-UV-400-30 \cite{ledrref})
 illumination.
 When using LED illumination, short (11~ns) pulses were used to obtain the PE spectrum, while long (16~ns) pulses were used to collect signals with magnitudes comparable to those expected from the CRV counters during operation.
\section{Radiation damage measurements\label{hms}}
The photoelectron spectrum, the response to a  LED pulse, the dark count rates at thresholds of up to 5.5~PE, and 
the MPPC current as a function of applied bias voltage (I-V scans) were measured for 40 (five sets of eight) non-irradiated MPPCs. 
Each set of eight MPPCs will henceforth be referred to as a "panel." 

One panel was never irradiated while the
other four panels were exposed to  different doses of 200~MeV protons (which emulate an equivalent  
1~MeV neutron dose under the NIEL approximation \cite{why200MeV}). 
The radiation doses were delivered with a 200~MeV proton beam at the Northwestern Proton Therapy Facility in Warrenville, IL  
with intensities in the range [5$\times 10^{9}$~p/cm$^2$, 5$\times 10^{10}$~p/cm$^2$]. The MPPCs were irradiated on their carrier board panels and were unbiased during irradiation as bulk damage is the primary physical process of interest \cite{why200MeV}. The dose uniformity across the area was measured on site at Warrenville and was within $\pm$5\%.

After irradiation, the performance measurements were repeated on all MPPC panels after accelerated annealing.   
Annealing was accelerated by holding the MPPCs at 60\degree C for 80 minutes, corresponding to about ten days of room temperature annealing.  
%
\subsection{\label{iv_scans}I-V scans}
\begin{figure}[hp]
\centering
{
\includegraphics[scale=0.40]{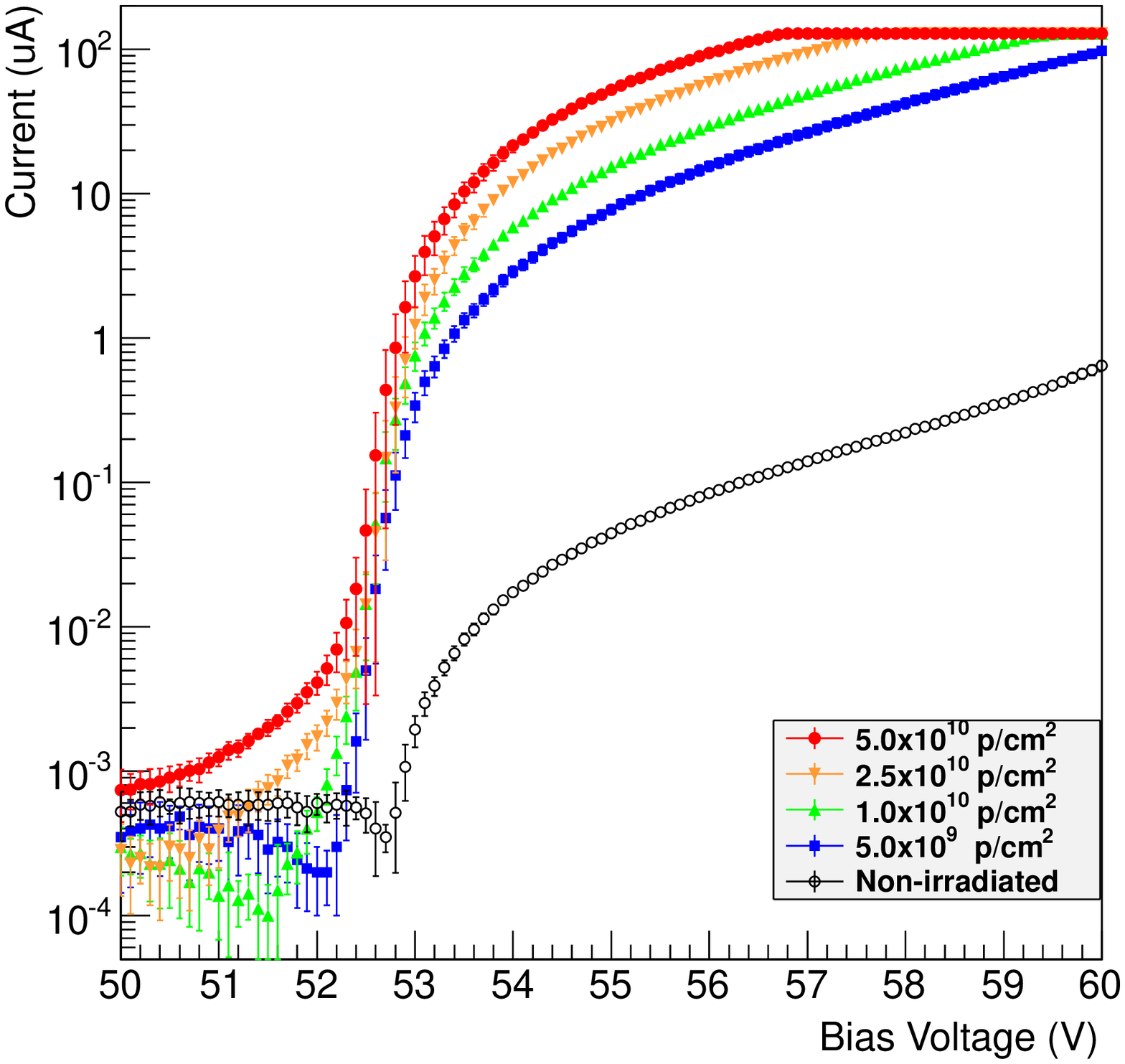}
\includegraphics[scale=0.40]{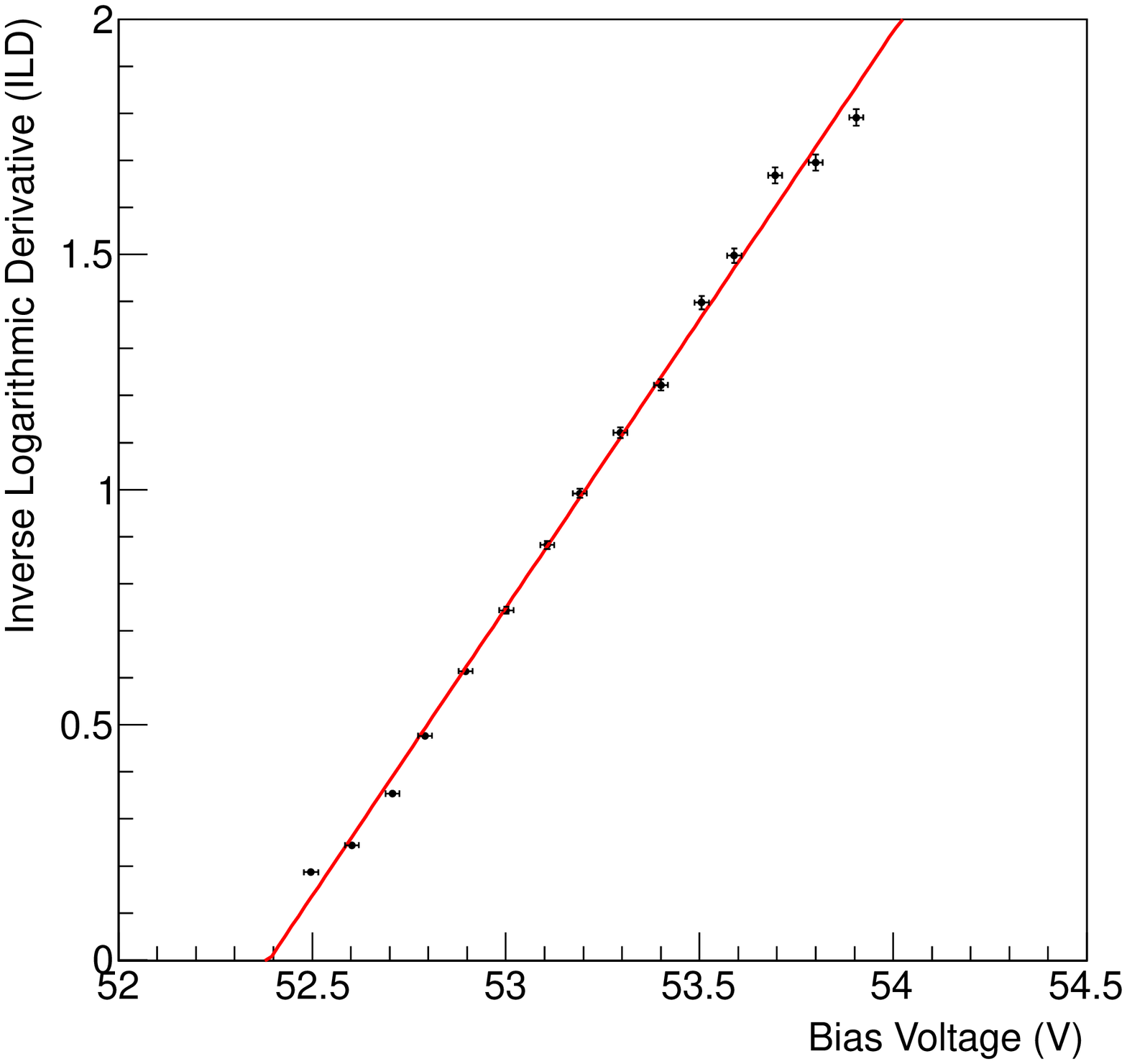}
}
\leftline{ \hspace{3.7cm}{\bf (a)} \hfill\hspace{3.7cm} {\bf (b)} \hfill}
\caption{\label{fig:2x2_IV_pk21_pk22}(a) I-V scans before irradiation (black empty circles) 
and after irradiation (filled shapes), for the four irradiated panels.  
Points correspond to the mean MPPC current and uncertainties are the spread of the MPPC current;  (b) the ILD near the breakdown region, for an individual irradiated MPPC.  The extrapolation of the ILD to zero is taken as the breakdown voltage. }
\centering
 {
 \includegraphics[scale=0.40]{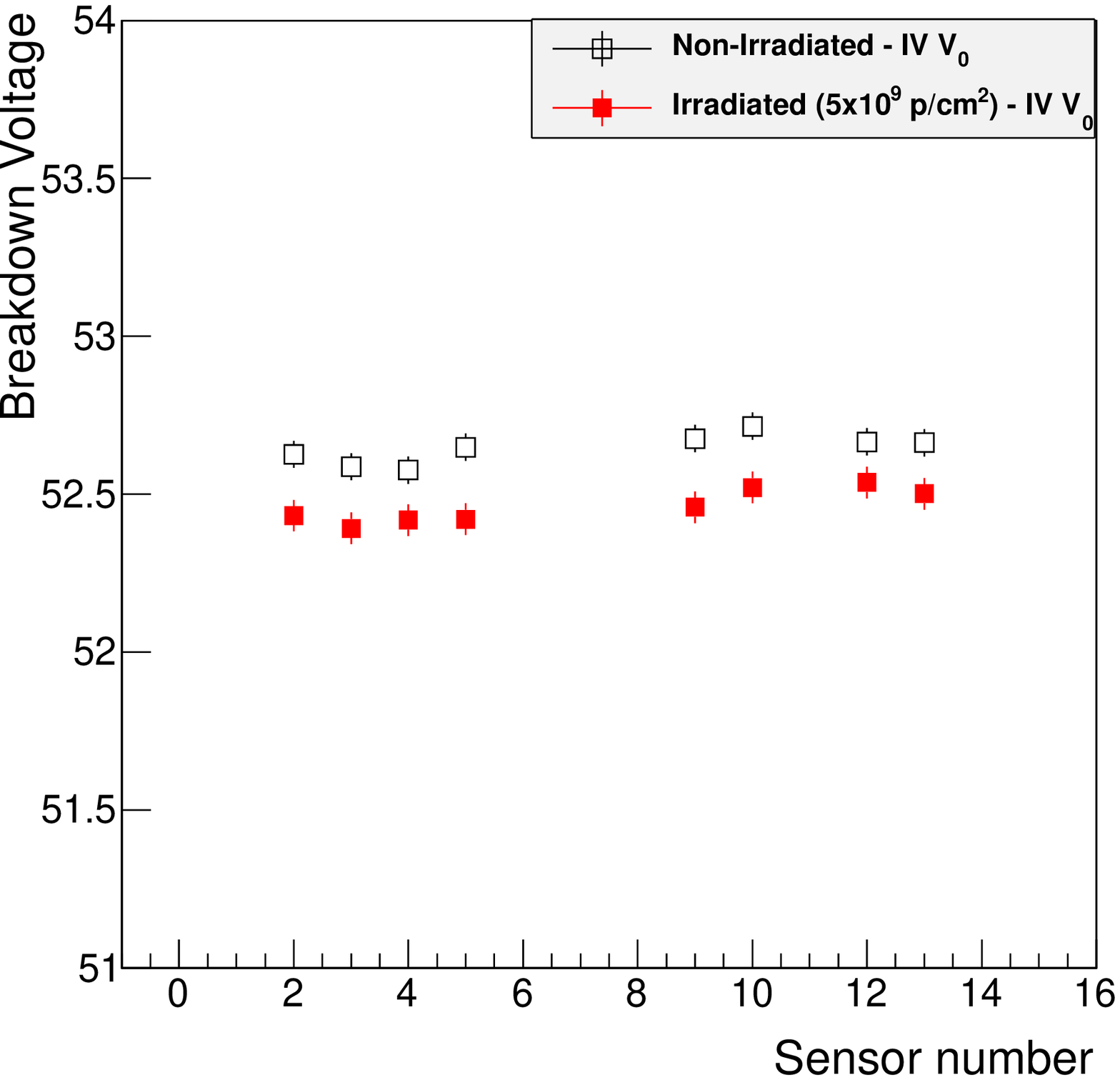}
 \includegraphics[scale=0.40]{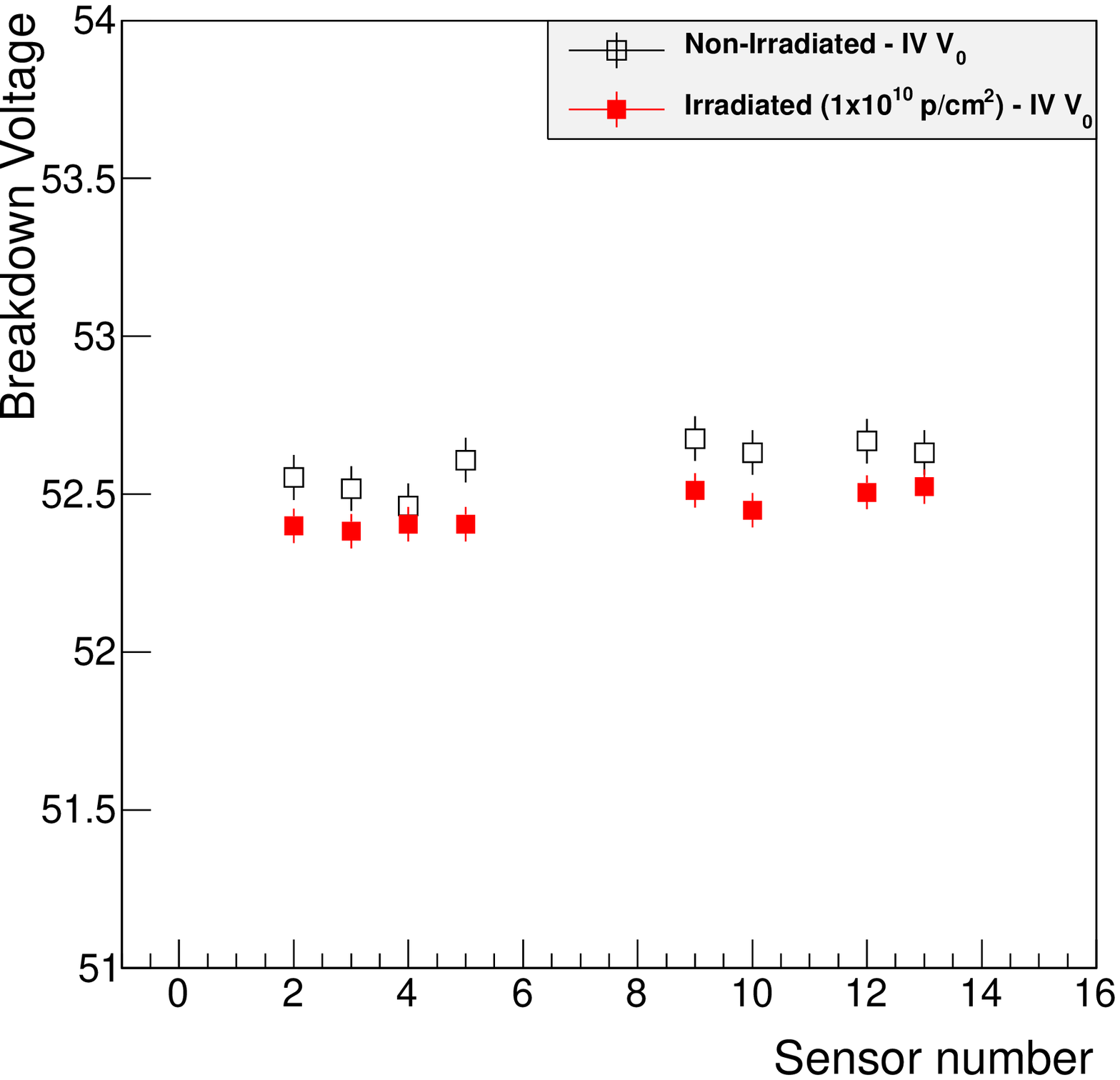} 
 }
 \leftline{ \hspace{3.7cm}{\bf (a)} \hfill\hspace{3.7cm} {\bf (b)} \hfill}
  \caption{\label{fig:2x2_ivBreakdown}   
Plots of the MPPC breakdown voltage per MPPC number calculated using the ILD method. Black empty squares denote the non-irradiated MPPCs, while red filled squares denote irradiated MPPCs at (a) 5$\times 10^{9}$ p/cm$^2$ and (b) 1$\times 10^{10}$ p/cm$^2$. 
}
\end{figure}

Using the FEB, we measured the MPPC currents as a function of applied bias voltage for each panel. 
These measurements were done before and after irradiation of the panels, as shown in Fig.~\ref{fig:2x2_IV_pk21_pk22}a, to determine any change in breakdown voltage and current levels.  

 The breakdown voltage can be determined using the inverse logarithmic derivative (ILD) of the current with respect to the voltage taken from an I-V scan \cite{ILD}. 
Breakdown, or turn-on, occurs at the voltage where the derivative rapidly changes from zero to a large value or, equivalently, where the inverse of the derivative approaches zero. As the definition of the breakdown voltage, we use the extrapolation of the ILD to zero \cite{klanner1}.
Fig.~\ref{fig:2x2_IV_pk21_pk22}b shows the ILD as a function of bias voltage for an irradiated MPPC. 
 The I-V breakdown voltages, before and after irradiation, are compared in Fig.~\ref{fig:2x2_ivBreakdown} and are stable within 0.2~V. As shown in Fig.~\ref{fig:2x2_ivBreakdown}a and Fig.~\ref{fig:2x2_ivBreakdown}b the standard deviation for each non-irradiated panel was roughly 60 mV, and 55mV for irradiated panels.
%
\subsection{Photoelectron (PE) spectrum analysis}
\begin{figure}[ht]
\centering
 {
\includegraphics[scale=0.37]{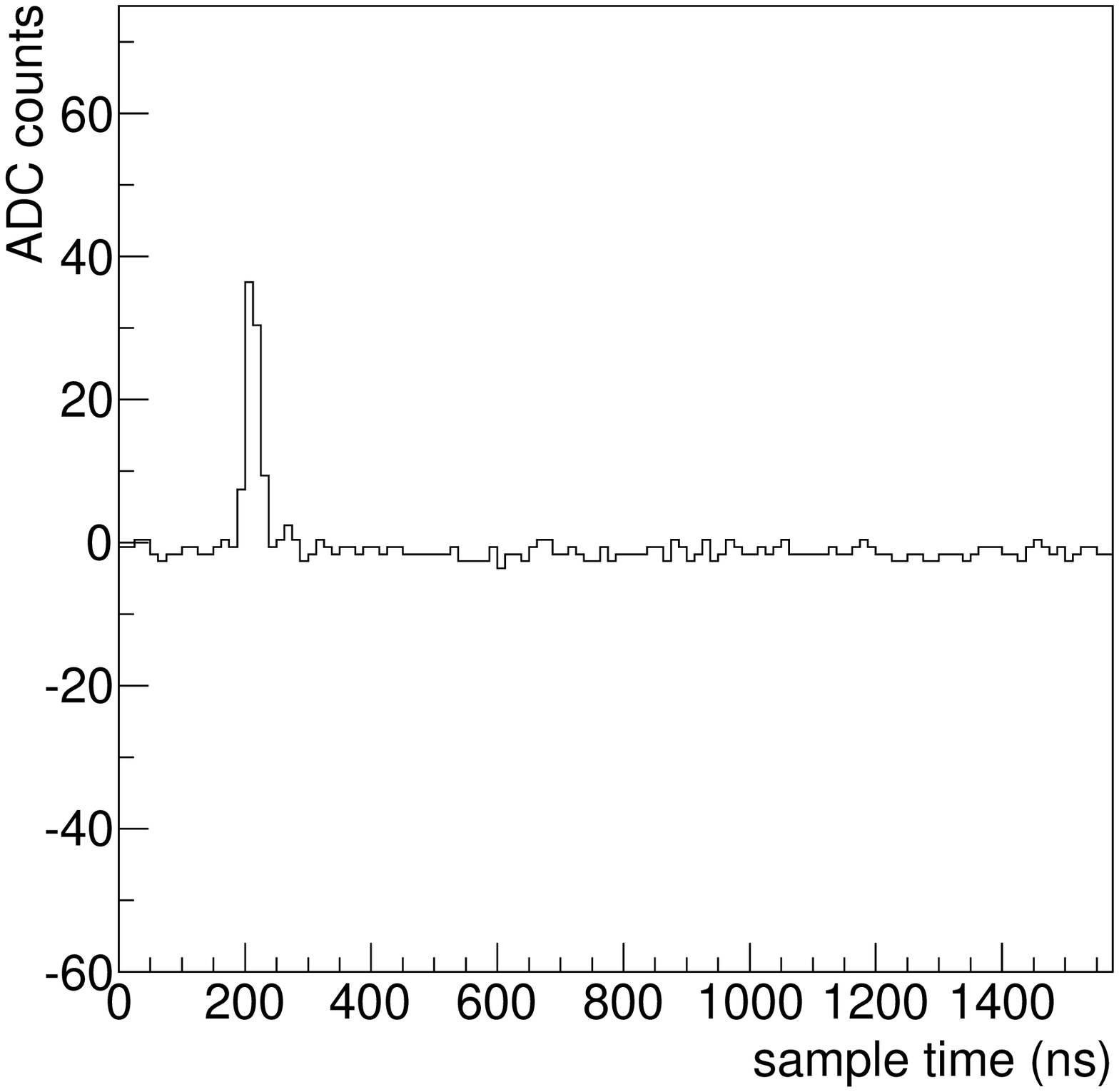}
\includegraphics[scale=0.37]{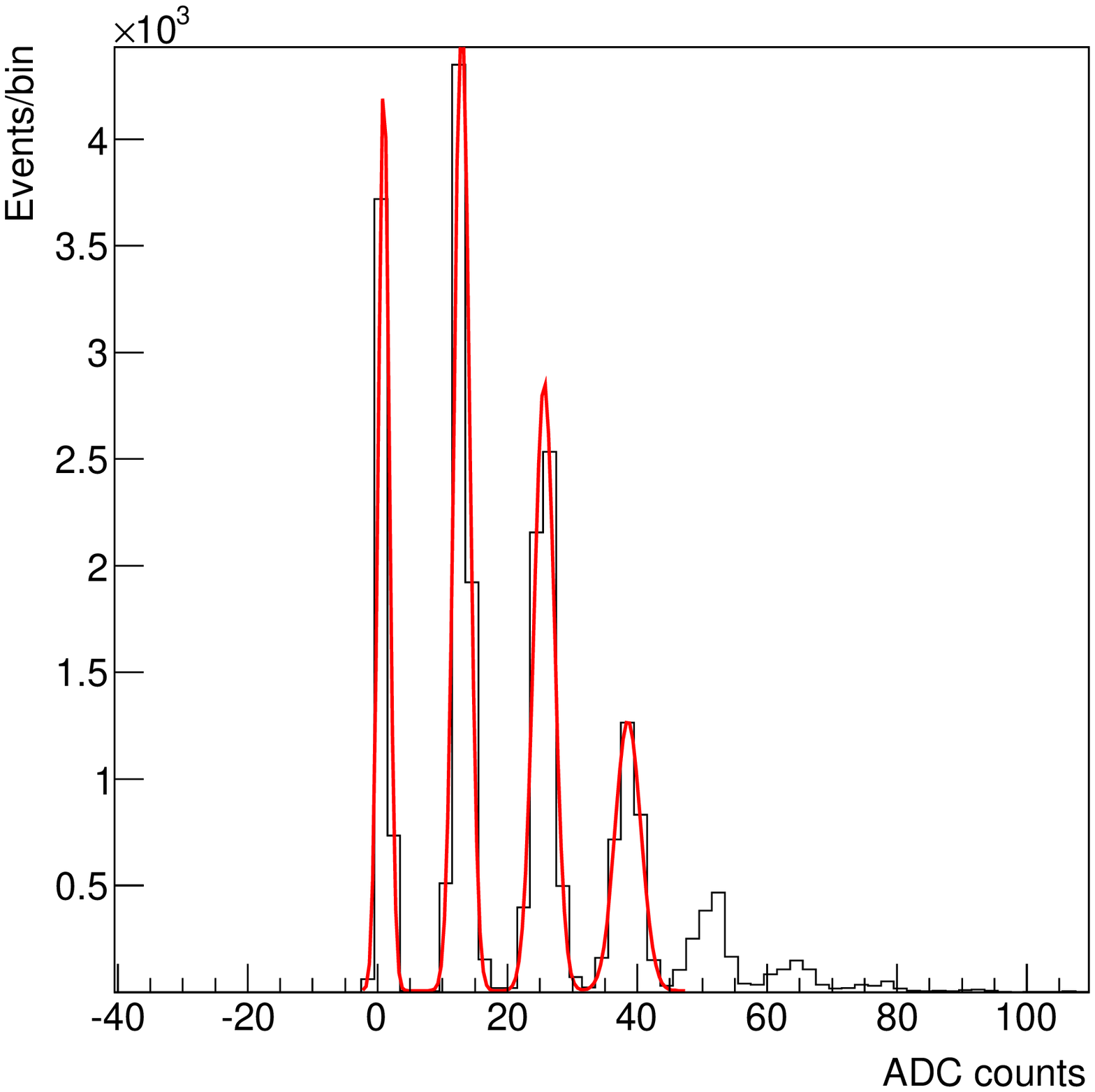}
 }
\leftline{ \hspace{3.7cm}{\bf (a)} \hfill\hspace{3.7cm} {\bf (b)} \hfill} 
\caption{\label{fig:2.0x2.0_clean} 
 (a) The non-irradiated LED signal waveform with a 12.5 ns time period for each sample bin, and (b) the non-irradiated LED photoelectron spectrum, with Gaussian fits over the peaks. 
}
\end{figure}

The FEB digitizes the MPPC analog signals using a 12-bit ADC with a 
12.5 ns sampling period. %
The FEB continuously samples the MPPC signal; for this study 128 samples of the digitized waveform were recorded as an event (see Fig.~\ref{fig:2.0x2.0_clean}a). The FEB is triggered by the LED pulser.  Both the time at which the waveform is initiated relative to the pulser signal and the waveform length are variable features of the FEB.
For every MPPC tested, we collected ~5,000-10,000 events at different bias voltages ($V_b$) in the range between 
the MPPC breakdown voltage ($V_{0}$)  and up to approximately 3~V over-voltage ($V_+$), where over-voltage is defined as: $V_{+}=V_{b}-V_{0}$ . The MPPCs are temperature sensitive, with up to a 50~mV/\degree C change in the breakdown voltage. Tests of the MPPCs are corrected according to this value. Ambient temperatures during MPPC data taking averaged 20.7~$^{\circ}$C and varied no more than 0.5~$^{\circ}$C.

Four panels were irradiated at the four different levels of 5$\times 10^{9}$~p/cm$^2$,  1$\times 10^{10}$~p/cm$^2$, 2.5$\times 10^{10}$ p/cm$^2$, and 5$\times 10^{10}$~p/cm$^2$ all with $\pm$5\% dose uniformity across the panel area. The two lowest doses are the primary focus due to a lack of photoelectron peak resolution at the higher doses. Only a few percent of the MPPCs for Mu2e will be exposed to a fluence above 5$\times 10^{9}$~n/cm$^2$.

To obtain the photoelectron spectrum, a 11 ns pulse is used to trigger both the LED and FEB. Figure~\ref{fig:2.0x2.0_clean}a shows a FEB waveform captured when a MPPC is illuminated with the LED.  The signal maximum in a five-sample region centered on the arrival time of the MPPC signal is taken as the MPPC response.  To find pedestals, we histogrammed samples selected from the waveform at a time interval preceding the LED pulse and fit the distribution with a Gaussian curve.  The mean of the fit or the pedestal is then subtracted from the maximum signals.  Figure~\ref{fig:2.0x2.0_clean}b shows a non-irradiated MPPC PE spectrum where the first peak at zero ADC counts corresponds to the electronics pedestal.  We fit the PE spectrum with a series of Gaussian curves and calculate the gain of the MPPC as the difference between the mean values of the electronics pedestal and the first PE peak. 
\subsection{Device gain before and after irradiation}
An apparent reduction in gain after irradiation was observed. 
The observed gain reduction is due to an effective drop of the applied bias voltage.
Since irradiation increases MPPC current as shown in the I-V scans, there is a voltage drop across the net series resistance in the external circuit, effectively decreasing the applied bias voltage.
The voltage drop is estimated via the following procedure: for each MPPC the current is measured for each bias voltage step and plotted to form an I-V scan. Then the I-V current and external resistances on the external circuit (totals to 8.04 $k\Omega\;\pm 2\%$) are used to correct the voltage drop for each bias voltage in the gain plots. When this voltage drop is accounted for, the apparent behavior of the gain decreasing with increased bias voltage is eliminated. 
\begin{figure}[hp]
\centering
 {
 \includegraphics[scale=0.40]{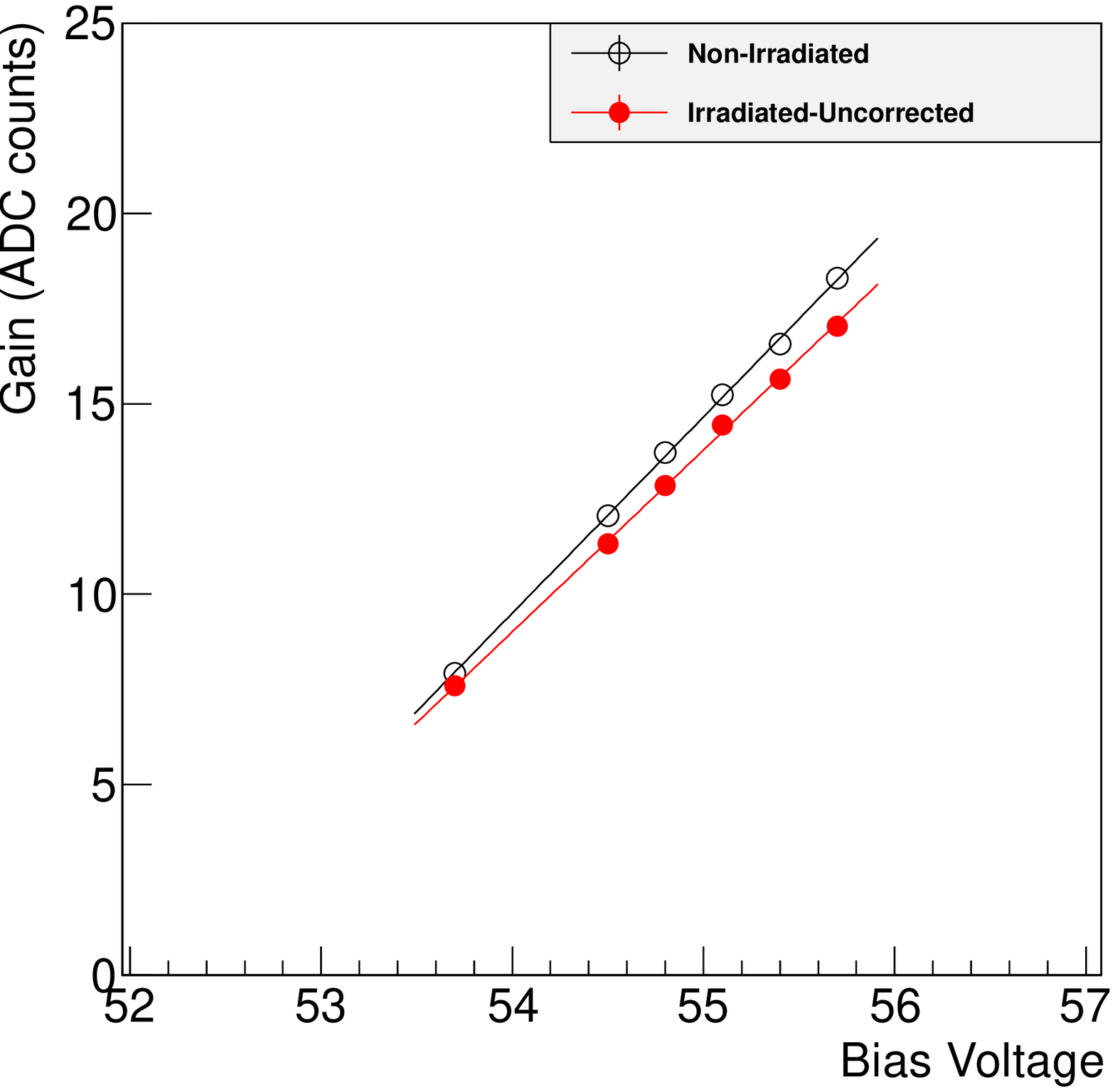}
 \includegraphics[scale=0.40]{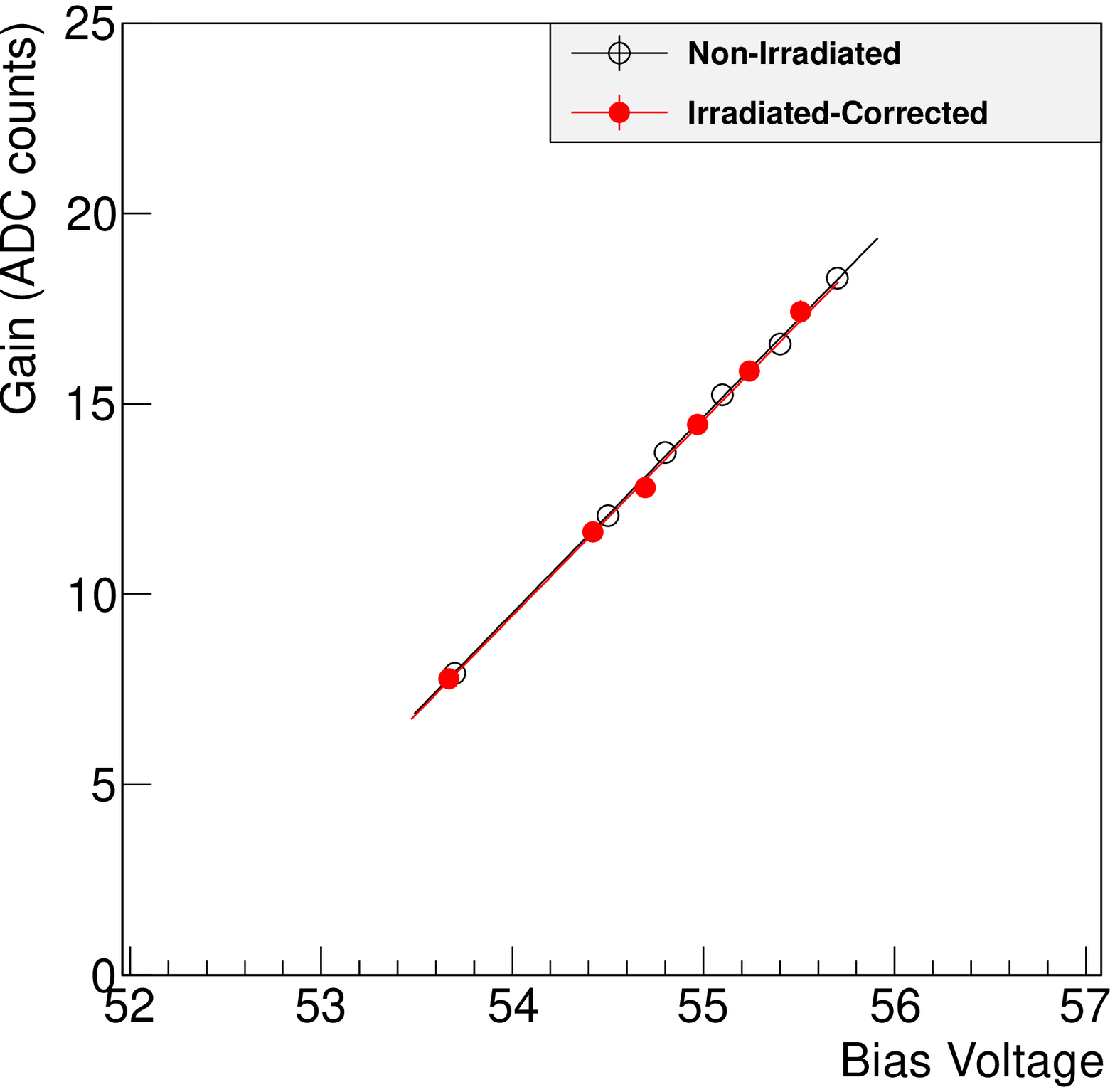}
 }
 \leftline{ \hspace{3.7cm}{\bf (a)} \hfill\hspace{3.7cm} {\bf (b)} \hfill}
  \caption{\label{fig:voltage_drop}   
 Effect of external resistor voltage drop on the gains for MPPCs irradiated with 1$\times 10^{10}$ p/cm$^2$.  
 The black empty circles in both plots show the gain before irradiation. 
 The red filled circles in the left plot show the gains after irradiation at the uncorrected bias voltage while the red filled circles in the right plot show the gains at the corrected bias after irradiation. 
}
 {
 \includegraphics[scale=0.40]{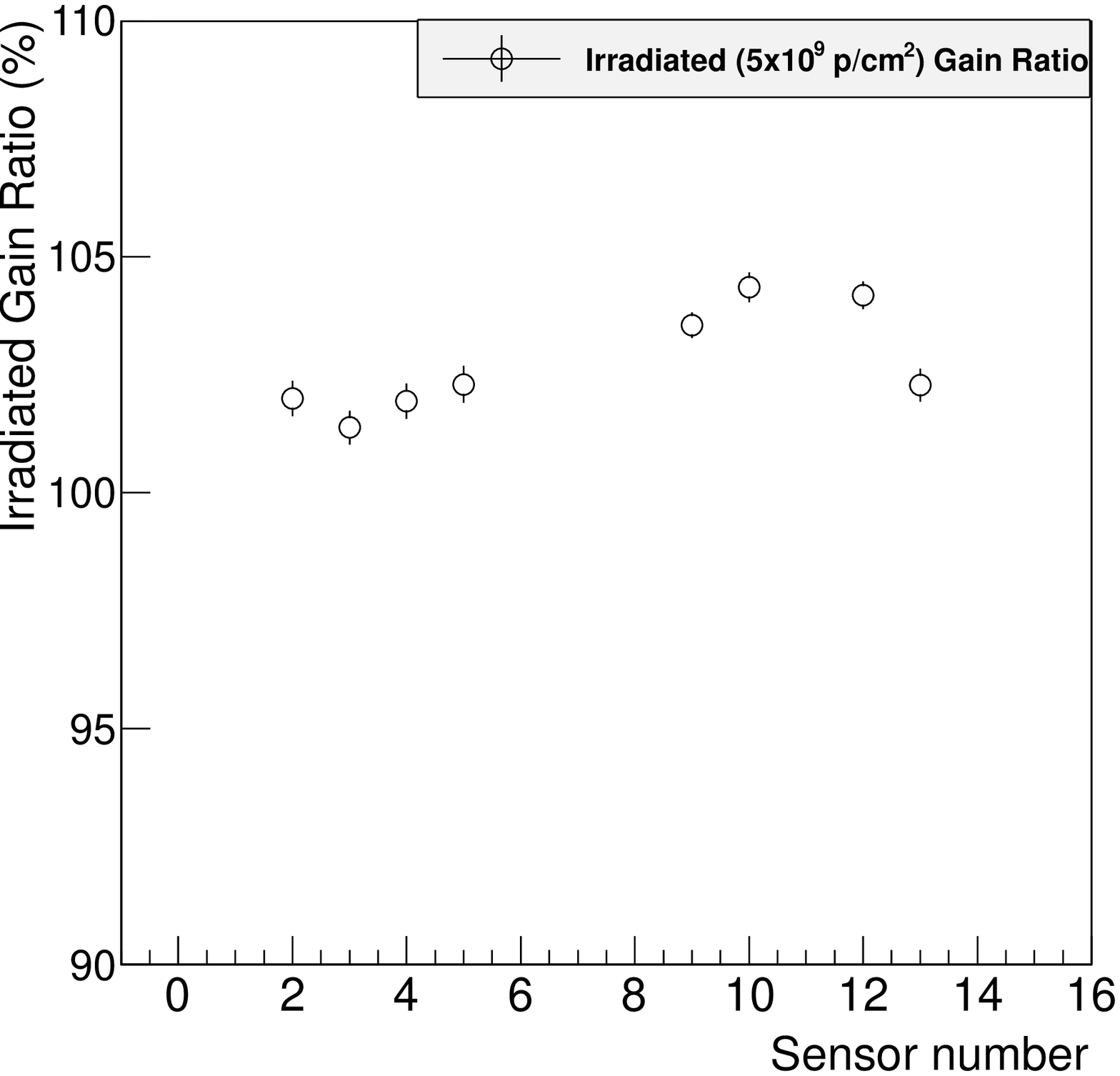}
 \includegraphics[scale=0.40]{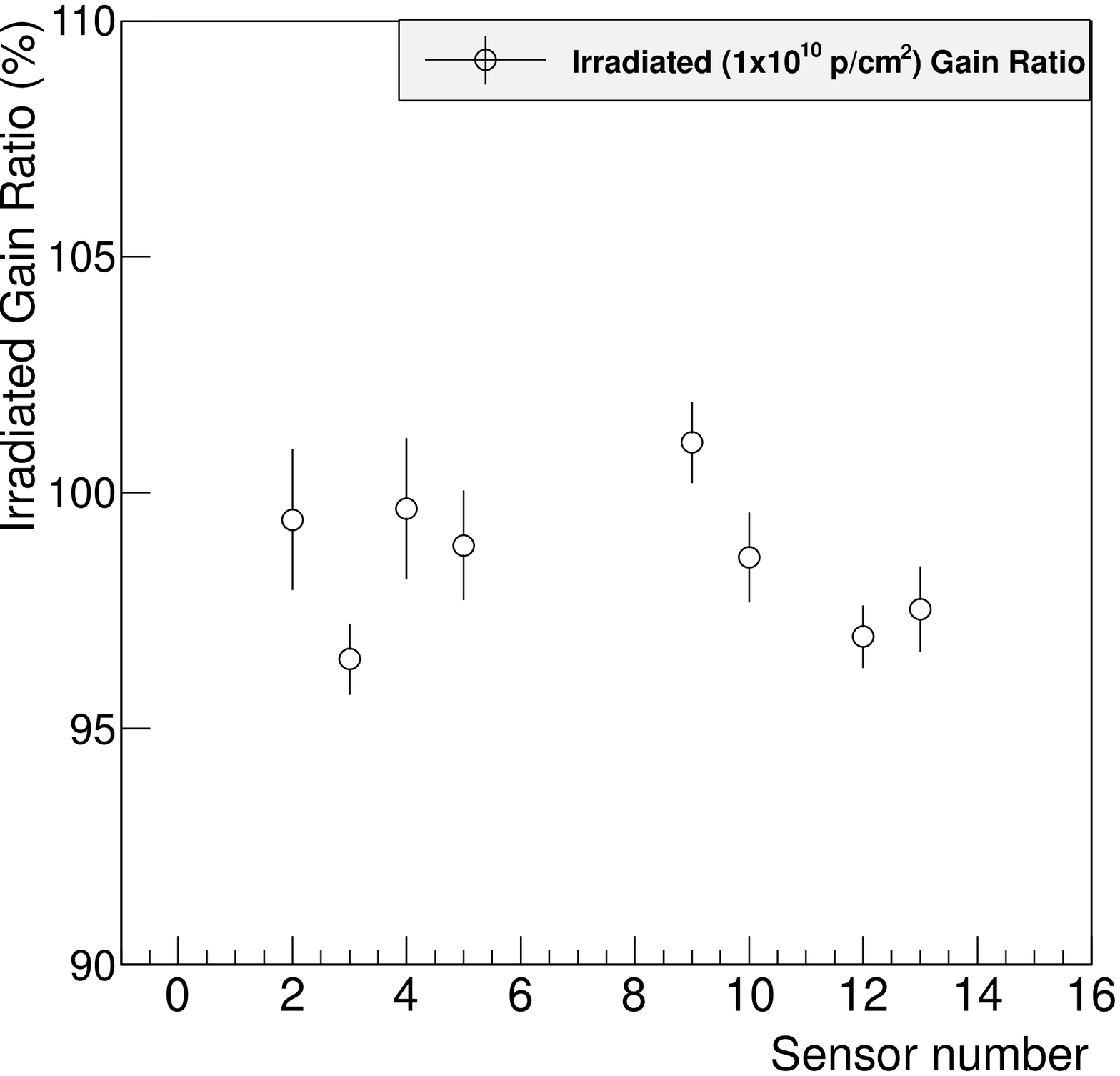}
 }
 \leftline{ \hspace{3.7cm}{\bf (a)} \hfill\hspace{3.7cm} {\bf (b)} \hfill}
  \caption{\label{fig:2x2_gain_adjusted}   
Plot of the MPPC gain ratio in percent, irradiated to non-irradiated, as a function of the MPPC number at V$_{0}$+2.3~V for (a) 5$\times 10^{9}$ p/cm$^2$ and (b) 1$\times 10^{10}$ p/cm$^2$.
}
\end{figure}
Figure~\ref{fig:voltage_drop} shows the gain as a function of bias voltage before and after irradiation, without (Fig.~\ref{fig:voltage_drop}a)  
and with (Fig.~\ref{fig:voltage_drop}b) the bias voltage correction 
for the MPPCs irradiated with 1$\times 10^{10}$ p/cm$^2$.  

Figure~\ref{fig:2x2_gain_adjusted} shows the ratio of the gain for all eight MPPCs in two of the tested panels before and after irradiation. The irradiated MPPC gains are corrected for the current draw, as in Fig.~\ref{fig:voltage_drop}b, by comparing to the irradiated points at the same bias over-voltage, V$_{0}$+2.3~V. The error bars for the points include systematic uncertainties associated with extraction of the PE peaks and I-V current measurements and voltage measurements, with the majority of the uncertainty from fitting PE peaks. The gains are stable within 4\%. The standard deviation prior to irradiation was roughly 60 mV, and 75 mV after irradiation.
\subsection{Breakdown voltages}
\begin{figure}[h!]
\centering
 {
 \includegraphics[scale=0.40]{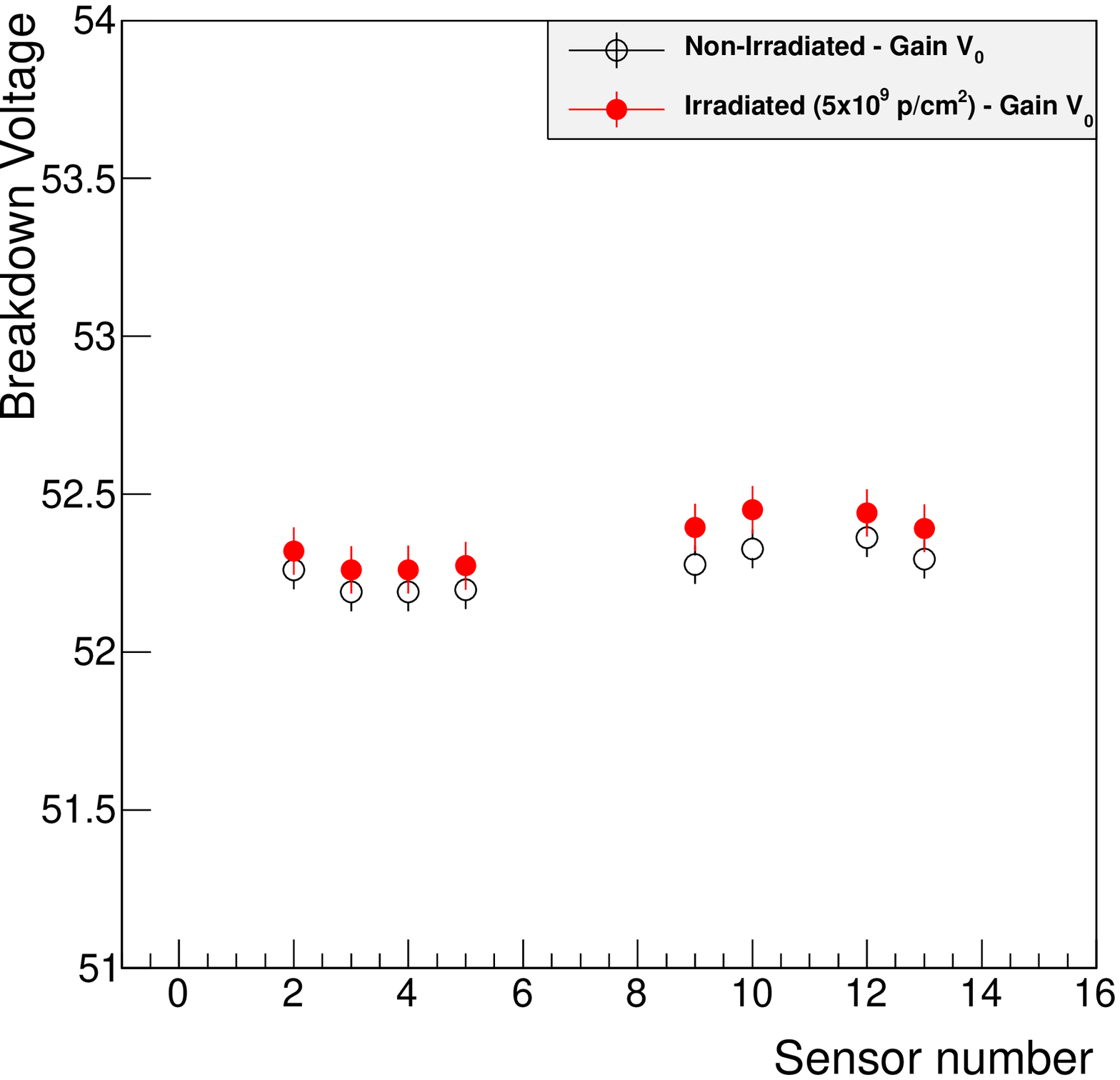}
 \includegraphics[scale=0.40]{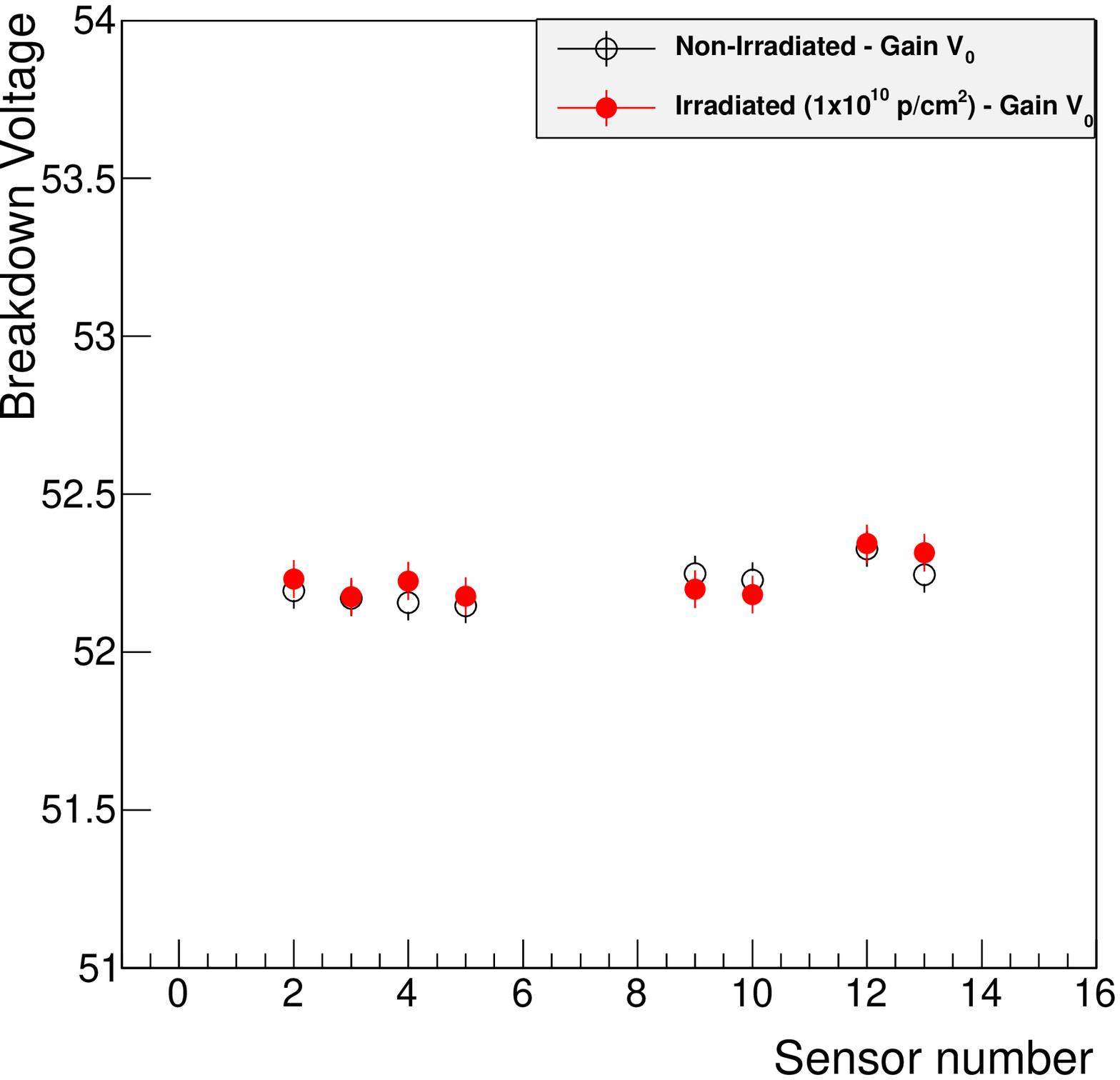} 
 }
 \leftline{ \hspace{3.7cm}{\bf (a)} \hfill\hspace{3.7cm} {\bf (b)} \hfill}
  \caption{\label{fig:2x2_gainBreakdown}   
MPPC breakdown voltage per MPPC number calculated from the gain versus bias voltage fits. Black empty circles denote the non-irradiated MPPCs, while red filled circles denote irradiated MPPCs at (a) 5$\times 10^{9}$ p/cm$^2$ and (b) 1$\times 10^{10}$ p/cm$^2$ . 
}
\end{figure}

We also measured the breakdown voltage from the gain plots by extrapolating the gain versus $V_b$ curve to zero gain (see Fig.~\ref{fig:voltage_drop}b). The non-irradiated and irradiated voltage breakdowns as determined with the gain are plotted per MPPC number in Fig.~\ref{fig:2x2_gainBreakdown}.   
Uncertainties in the breakdown plots correspond to the standard deviation of the mean breakdown voltage for each panel.  Overall, using the gain method, the average $V_0$ for the non-irradiated MPPCs is $52.2\pm0.05~V$; while the average $V_0$ for the irradiated MPPCs is $52.3\pm0.05~V$ for the 5$\times 10^{9}$~p/cm$^2$ panel and $52.4\pm0.06~V$ for the 1$\times 10^{10}$~p/cm$^2$ panel.
%
\begin{figure}[htp!]
\centering
 {
 \includegraphics[scale=0.40]{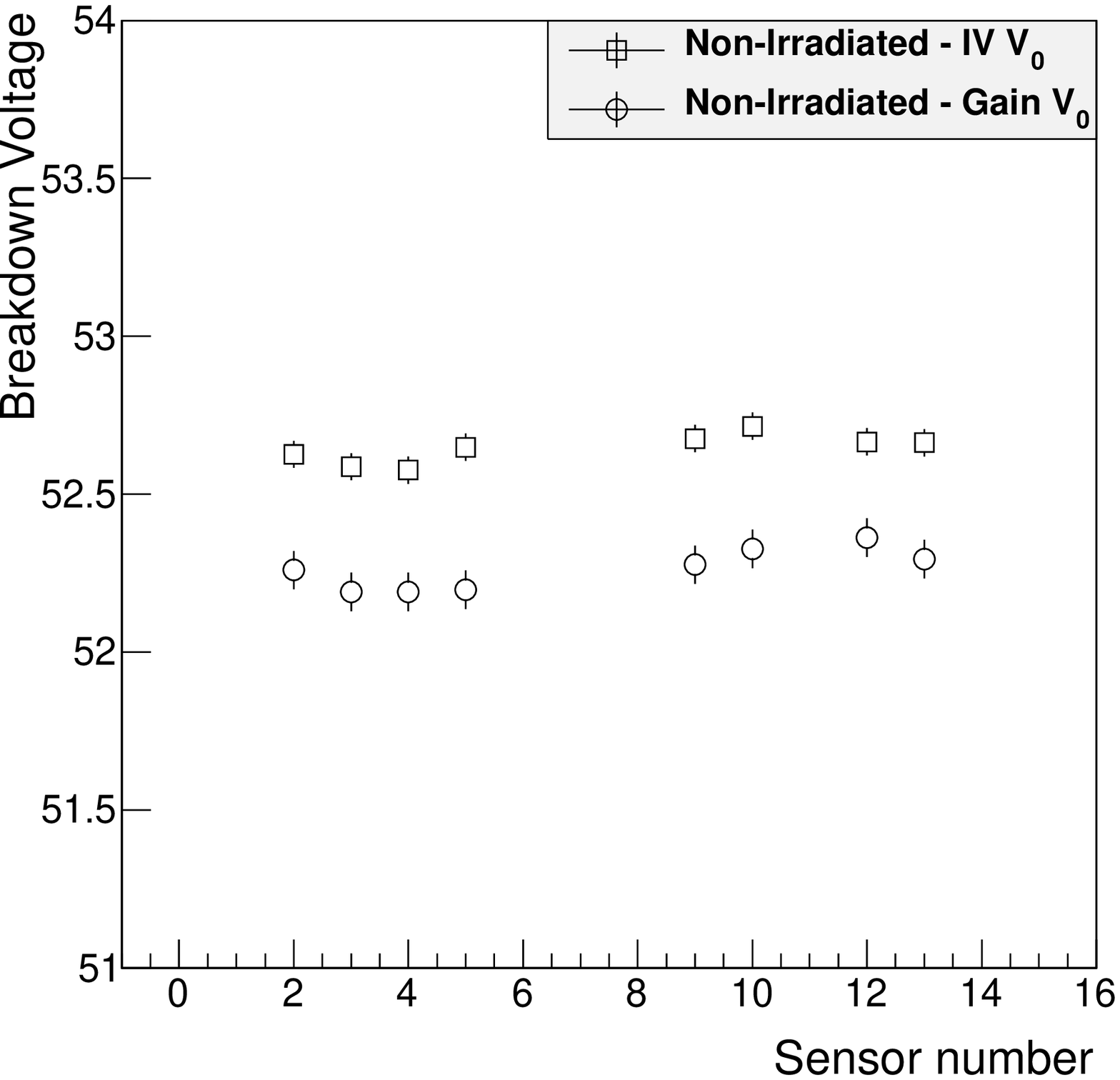}
 \includegraphics[scale=0.40]{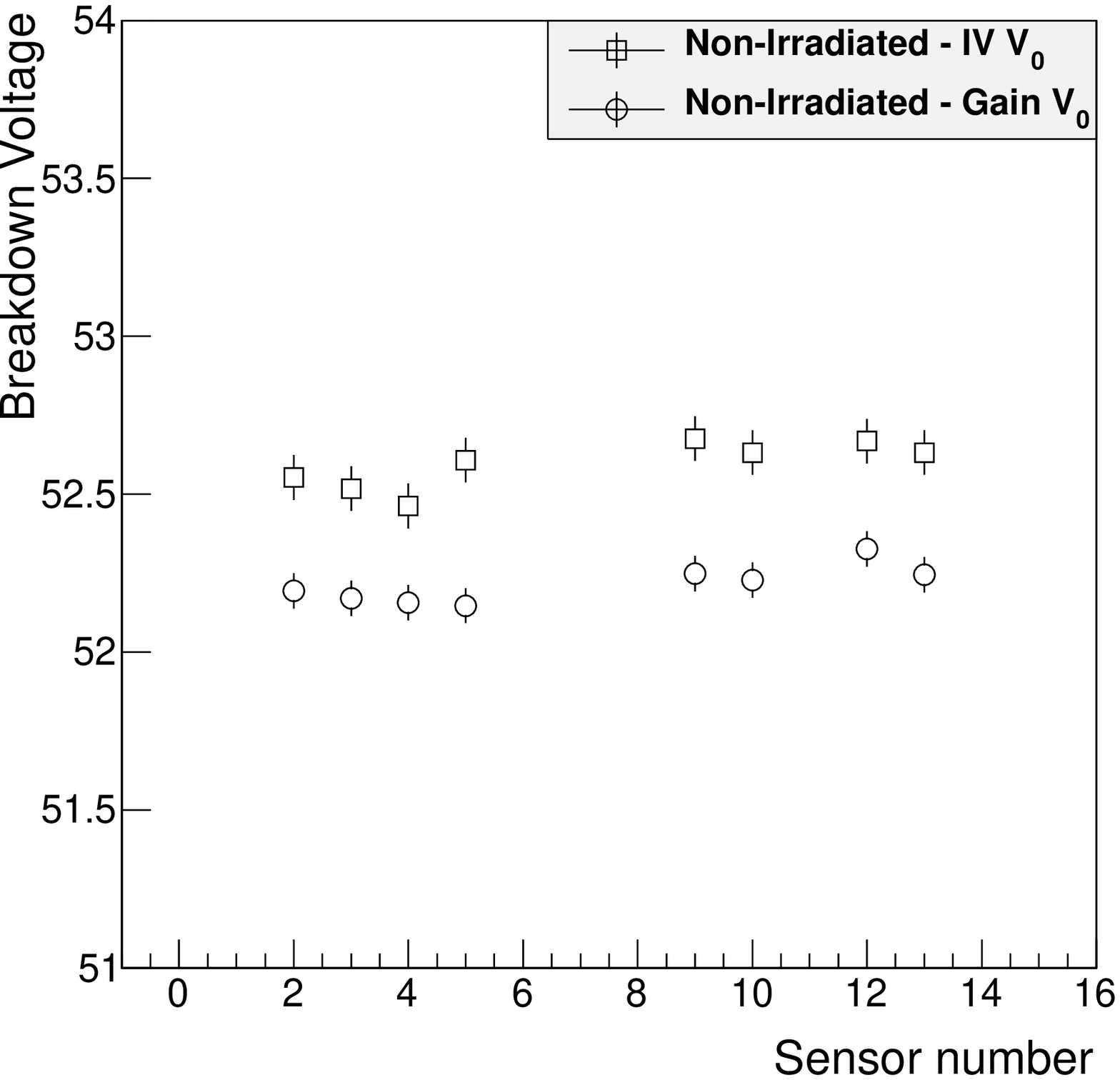} 
 }
 \leftline{ \hspace{4.0cm}{\bf (a)} \hfill\hspace{3.5cm} {\bf (b)} \hfill} 
\centering
 {
 \includegraphics[scale=0.40]{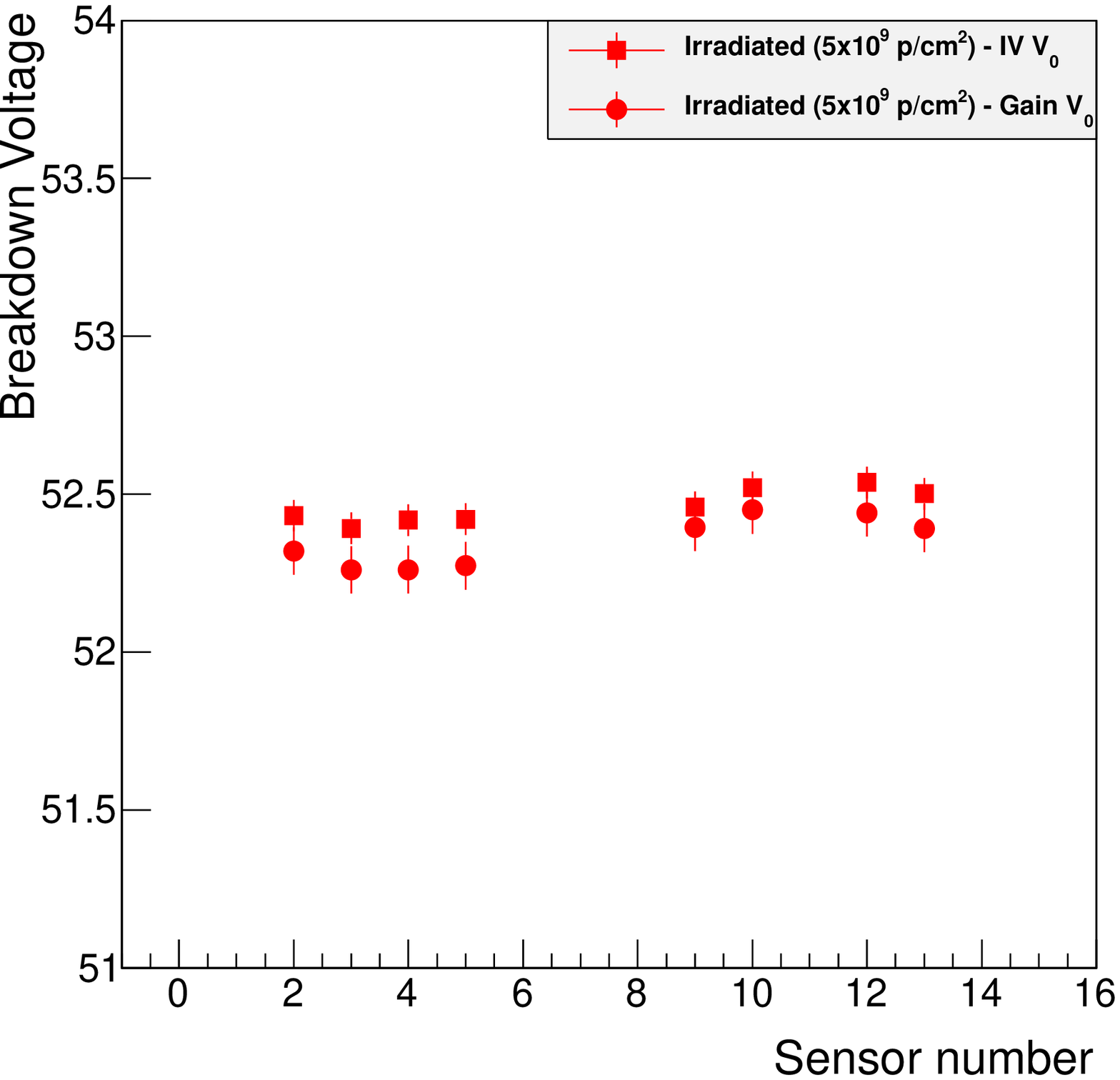}
 \includegraphics[scale=0.40]{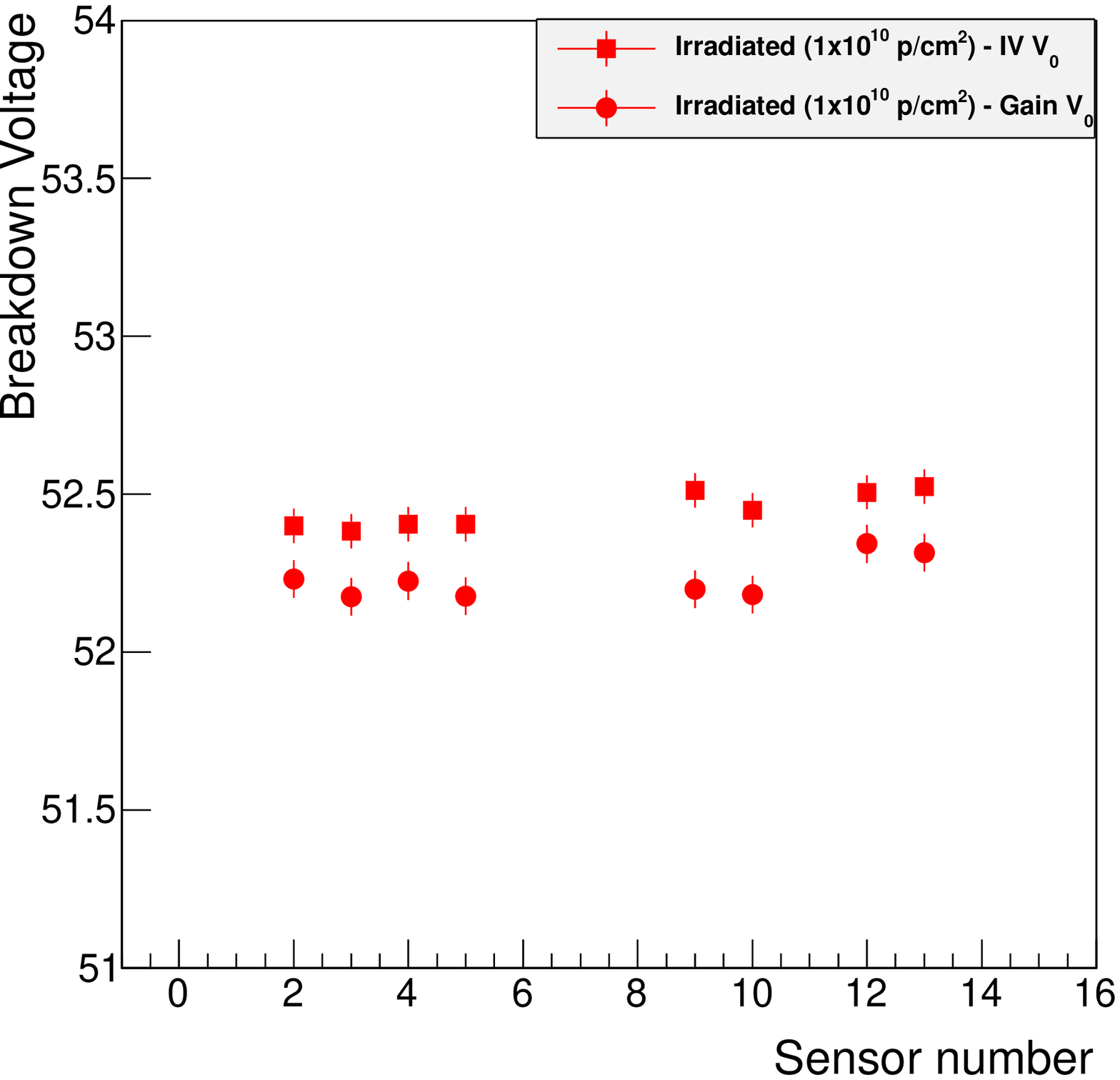} 
 }
 \leftline{ \hspace{4.0cm}{\bf (c)} \hfill\hspace{3.5cm} {\bf (d)} \hfill}
  \caption{\label{fig:2x2_gain-IV-br}   
Figures (a) and (b) show the breakdown voltages for the 5$\times 10^{9}$~p/cm$^2$ and the 1$\times 10^{10}$~p/cm$^{2}$ panels before irradiation using the I-V and gain methods, respectively. Figures (c) and (d) show the breakdown voltages for the two panels after irradiation.
}
\end{figure}

The two breakdown determination methods, gain and I-V, are compared in Fig.~\ref{fig:2x2_gain-IV-br}.  In general, breakdowns calculated with the I-V method are larger than those calculated with the gain method. This effect has been observed in other radiation damage studies \cite{klanner1}. Differences between the breakdown determination methods are likely a result of the turn-on (I-V) and turn-off (gain) breakdown state of the MPPCs \cite{MPF}.
\subsection{Dark count rates}

\begin{figure}[h!t]
\centering
 {
\includegraphics[scale=0.25]{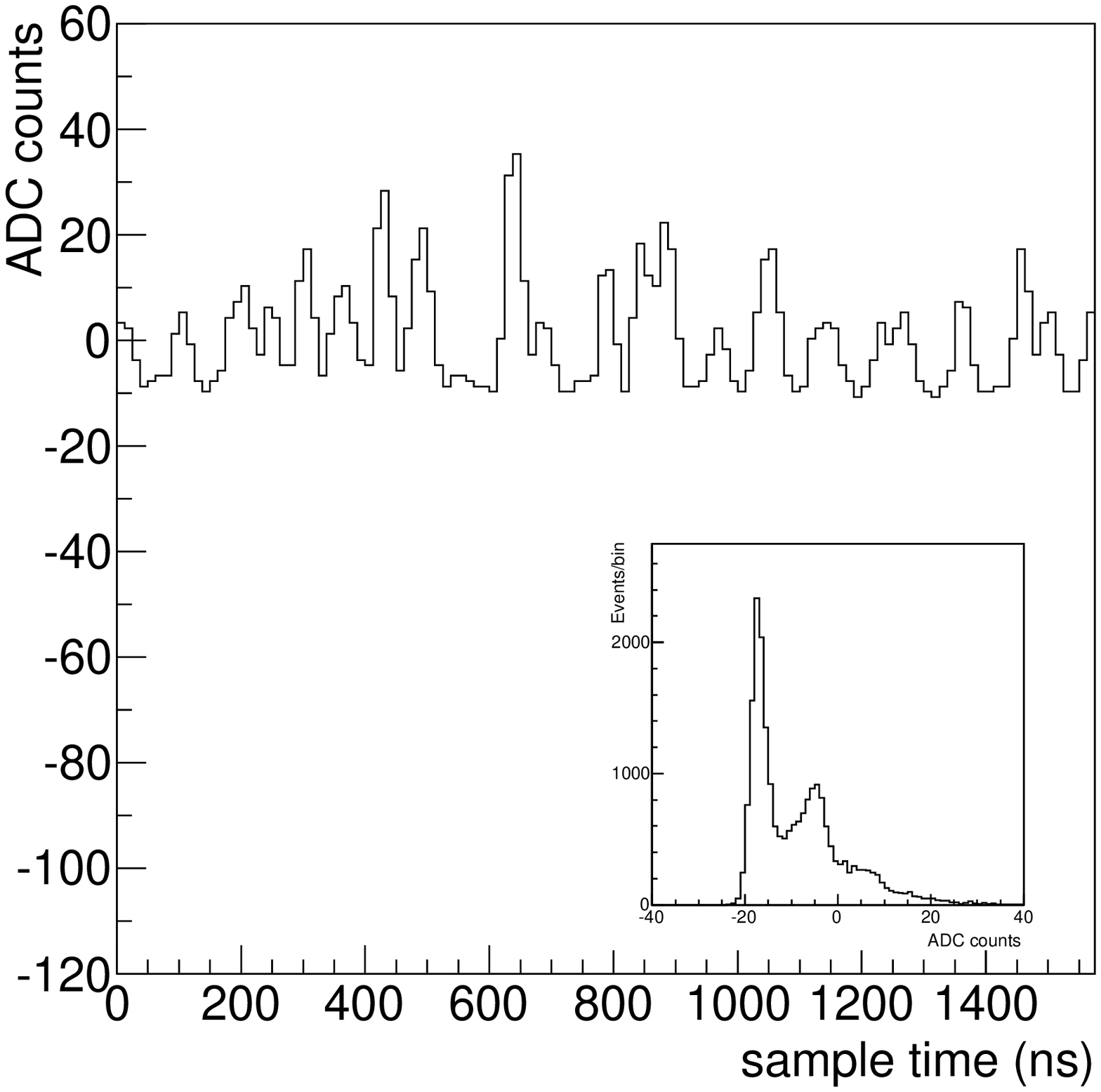}
\includegraphics[scale=0.25]{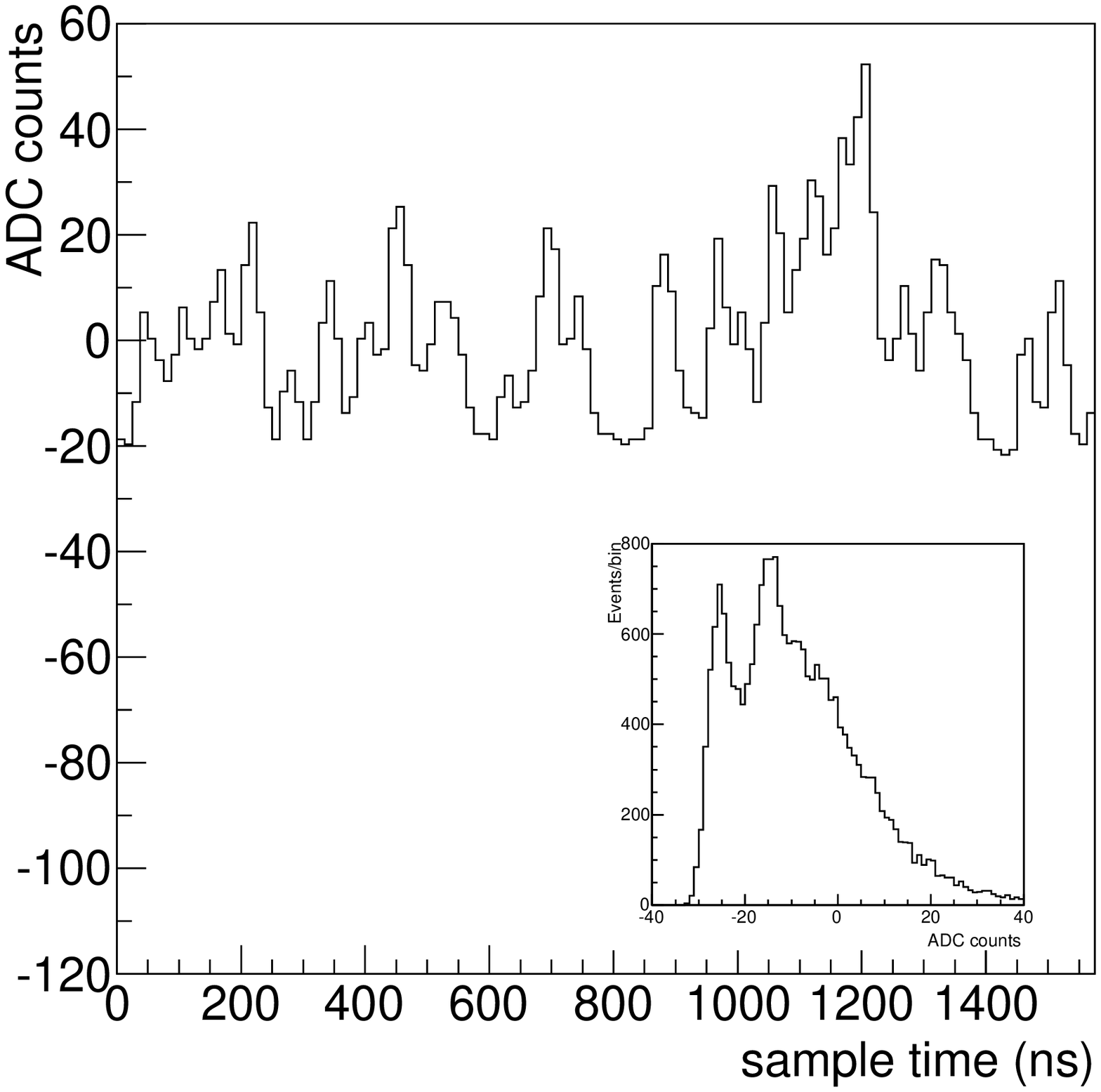}
\includegraphics[scale=0.25]{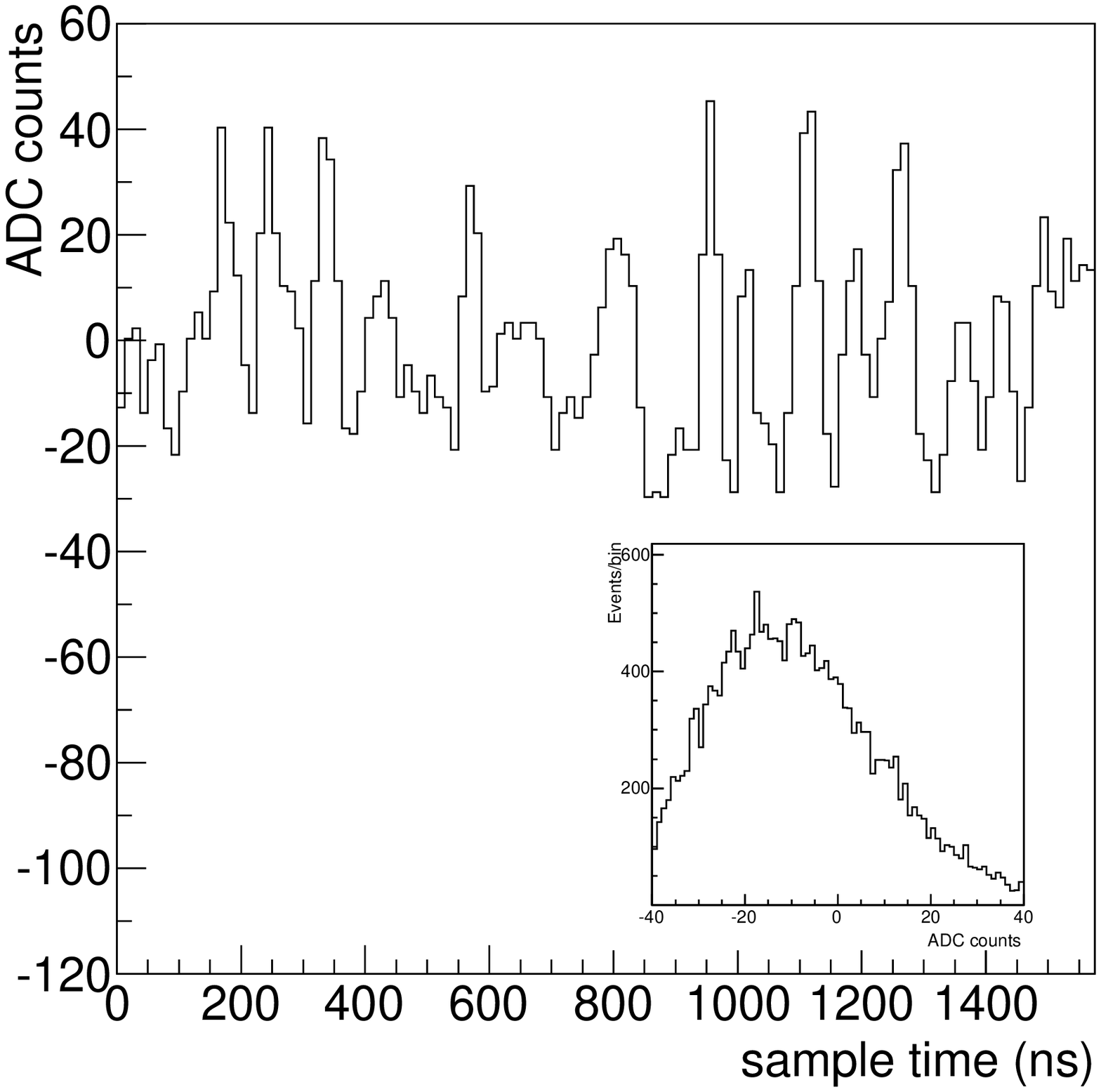}
 }
\leftline{ \hspace{2.4cm}{\bf (a)} \hfill\hspace{2.4cm} {\bf (b)} \hfill \hspace{2.4cm} {\bf (c)} \hfill}
\caption{\label{fig:2.0x2.0_wf} 
 Example waveforms with their respective PE spectrum (inset) which is measured outside of the 11~ns LED pulse region, obtained from MPPCs irradiated at 
 (a) 5$\times 10^{9}$~p/cm$^2$, 
 (b) 1$\times 10^{10}$~p/cm$^{2}$,  and   (c) 2.5$\times 10^{10}$~p/cm$^{2}$. 
 Data were taken at a bias overvoltage of V$_{0}$+2.3~V.
}
 {
\includegraphics[scale=0.25]{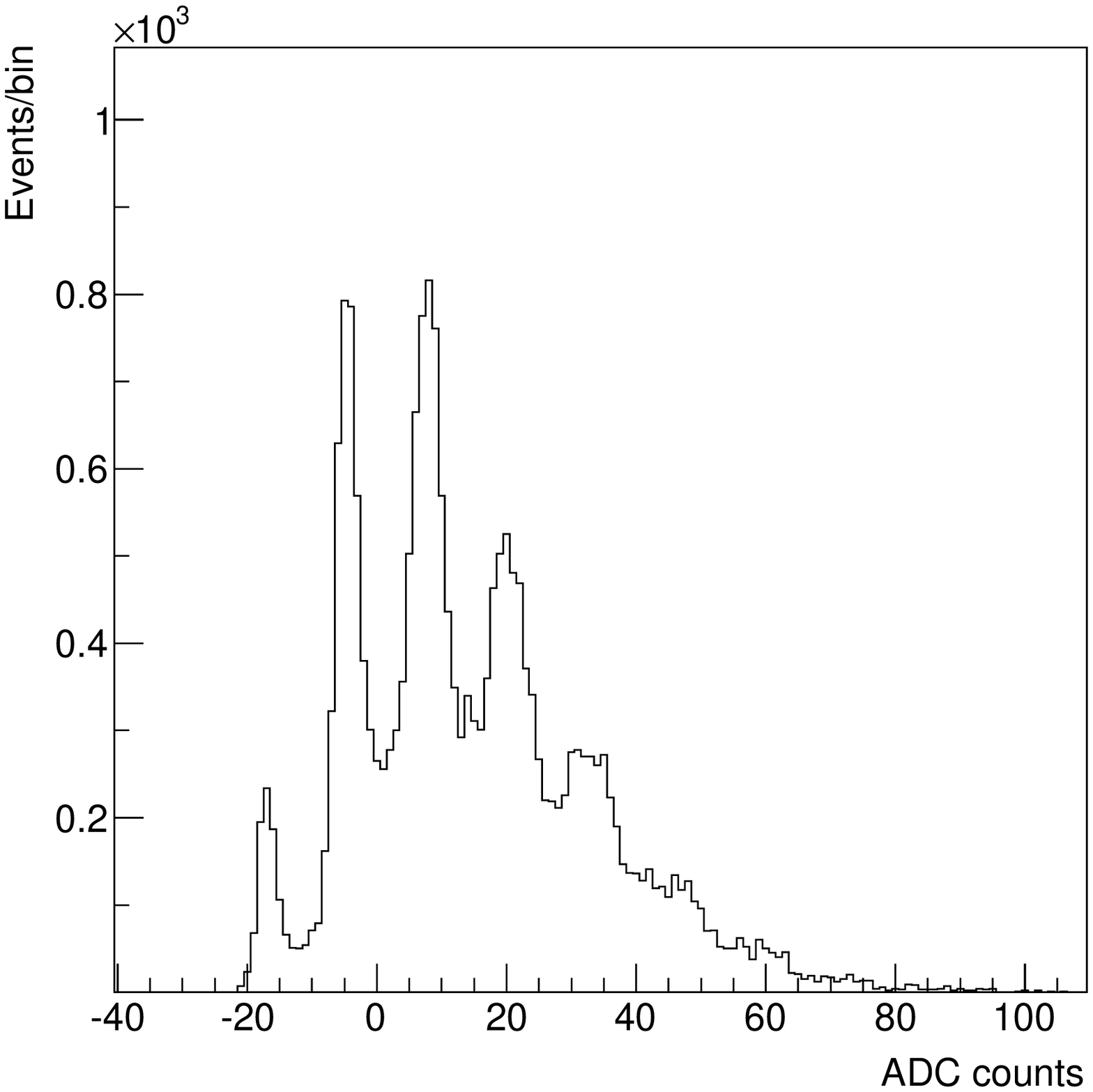}
\includegraphics[scale=0.25]{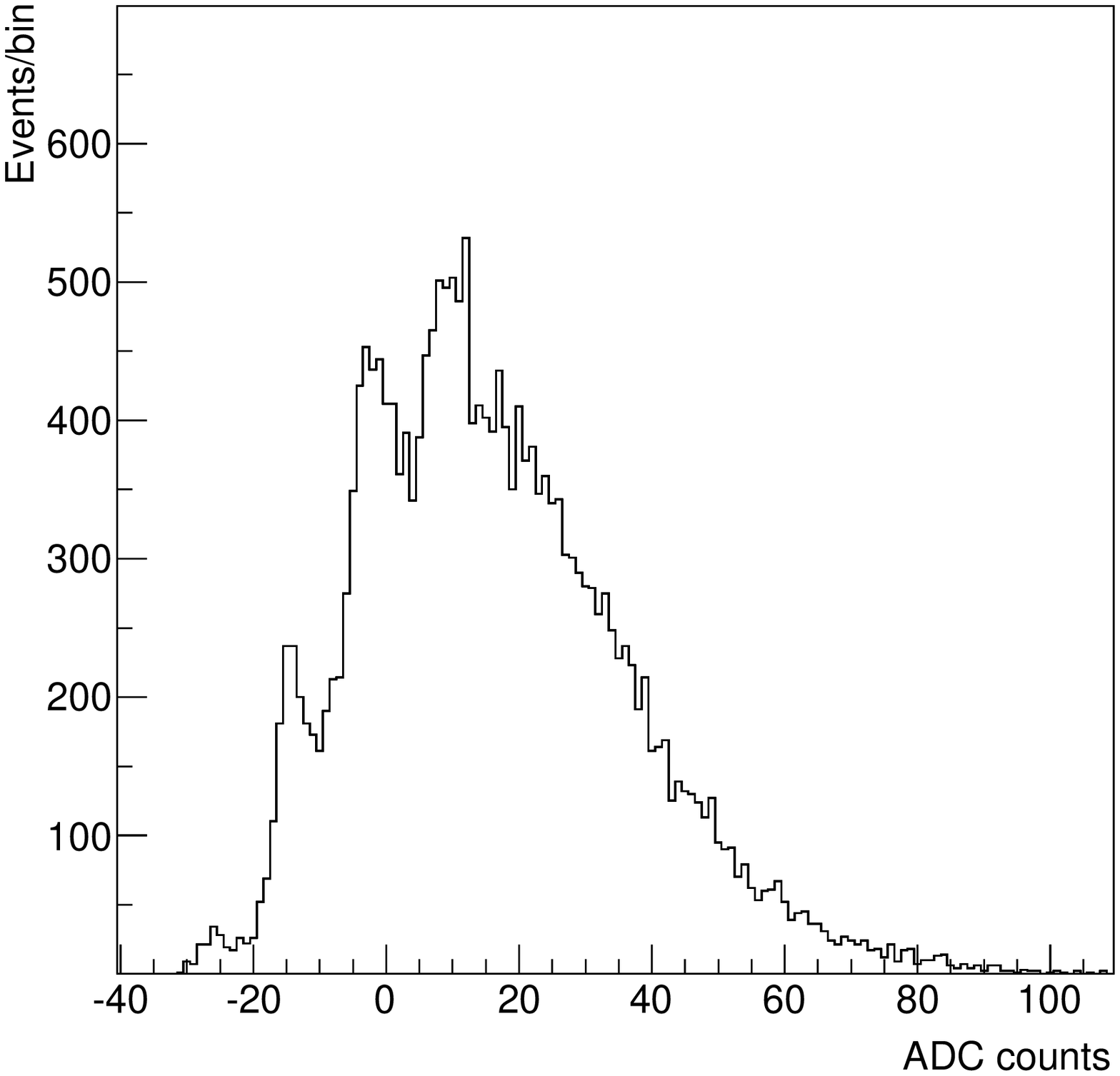}
\includegraphics[scale=0.25]{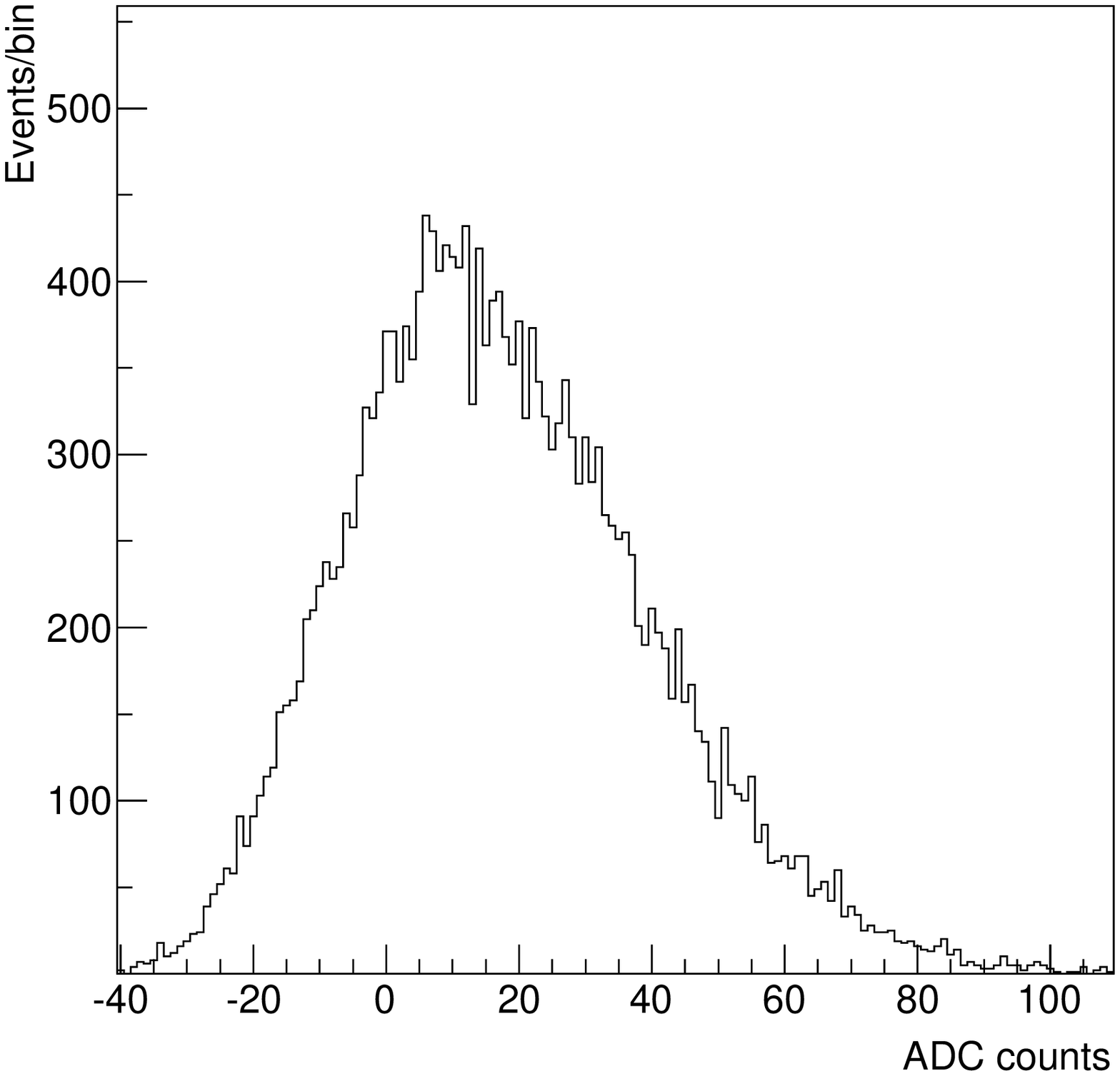}
 }
\leftline{ \hspace{2.3cm}{\bf (a)} \hfill\hspace{3.2cm} {\bf (b)} \hfill \hspace{3.2cm} {\bf (c)} \hfill}
\caption{\label{fig:2.0x2.0_sp} 
 PE spectra for the MPPCs, drawn from the 11~ns LED pulse region, irradiated at 
 (a) 5$\times 10^{9}$~p/cm$^2$,
 (b) 1$\times 10^{10}$~p/cm$^{2}$,  and   (c) 2.5$\times 10^{10}$~p/cm$^{2}$.  Data were taken at a  bias overvoltage of V$_{0}$+2.3~V.
}
\end{figure}

LED and noise  data samples for each panel of MPPCs were collected at bias voltages  in the range of [53.7, 55.7]~V, corresponding to an overvoltage above breakdown in the range  [1.4, 3.4]~V.  
Figure~\ref{fig:2.0x2.0_wf} shows the waveforms and PE spectra, measured away from the LED pulse region. For irradiated MPPCs an increase in noise with fluence is evident. Figure~\ref{fig:2.0x2.0_sp}
shows the pedestal-uncorrected photoelectron spectra, with 11ns LED illumination, for panels after irradiation at the lowest three doses.  At 2.5$\times10^{10}$ p/cm$^2$, the very high noise rate obscures PE peak separation.

\begin{figure}[h]
\centering
\includegraphics[scale=0.49]{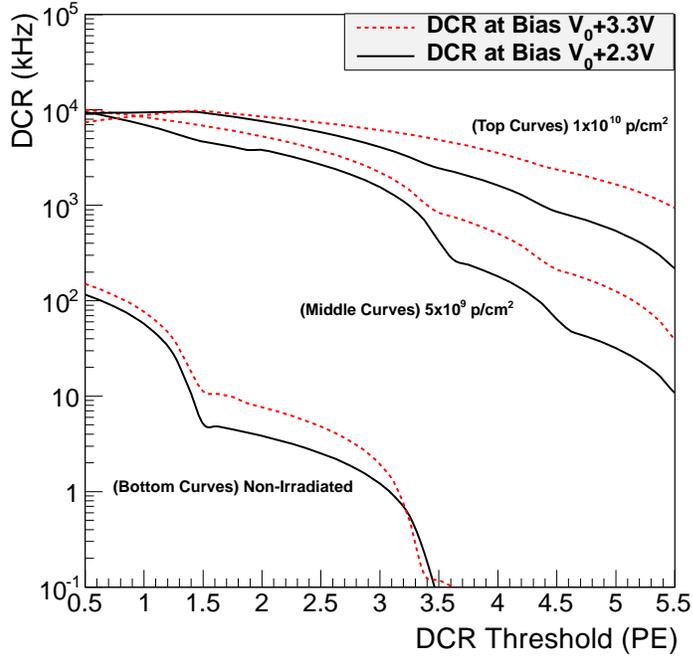}
\caption{\label{fig:2x2_dark_current} The dark count rates of the  MPPCs 
as a function of the PE threshold levels for two applied biases, V$_{0}$+2.3~V and V$_{0}$+3.3~V, for non-irradiated MPPCs (bottom); MPPCs irradiated at 5$\times 10^{9}$ p/cm$^2$ (middle); MPPCs irradiated at  1$\times 10^{10}$ p/cm$^2$ (top).}
\end{figure}

We define a noise peak (NP) as a contiguous set of samples in a waveform with all amplitudes above a threshold level, from a 0.5 PE threshold up to a 5.5 PE threshold.
The dark count rate (DCR) is then calculated as the frequency of these noise peaks in the analyzed waveforms:
\begin{equation}
\label{dcr}         
                             DCR  =  \sum NP/( 12.5~ns \times  \sum NS )~,
\end{equation}
where  $\sum NS$  is the total number of samples searched for the noise peaks, and 12.5~ns is the 
time between each sample measurement.
The mean DCRs for a set of eight MPPCs at a bias voltage of V$_{0}$+2.3~V
are shown in Table~\ref{tab:dcr_rates_2x2}.
Figure~\ref{fig:2x2_dark_current} shows the mean DCRs per panel
as a function of different threshold levels at two different applied biases, for the non-irradiated MPPCs and for MPPCs irradiated with  5$\times 10^{9}$ p/cm$^2$  and  1$\times 10^{10}$ p/cm$^2$. At the maximum fluence of 1$\times 10^{10}$ p/cm$^2$ 
the measured DCRs are below  $\sim$250~kHz for a 5.5~PE threshold. %
\begin{table}[h!]
\begin{center}
\vspace{10mm}
\caption{The mean DCR per panel at V$_b$=V$_{0}$+2.3~V bias voltage.} 
\begin{tabular}{|c|c|c|}
\hline                                    
 Irradiation  level (p/cm$^2$)  & Threshold (PE)       & DCR, kHz           \\
\hline
   No radiation                  &      0.5            &     124$\pm$4     \\
   No radiation                  &      3.5            &         $<$0.1       \\
   5$\times$10$^9$               &      3.5            &    284$\pm$58    \\
    5$\times$10$^9$              &     5.5             &    8.5$\pm$2.4   \\
   1$\times$10$^{10}$            &     3.5             &    2550$\pm$200   \\
   1$\times$10$^{10}$            &     5.5             &    260 $\pm$45    \\
\hline 
\end{tabular}
\label{tab:dcr_rates_2x2}
\vspace{10mm}
\end{center}
\end{table}
\pagebreak
\subsection{MPPC Response}
\begin{figure}[ht]
\centering
{
\includegraphics[scale=0.27]{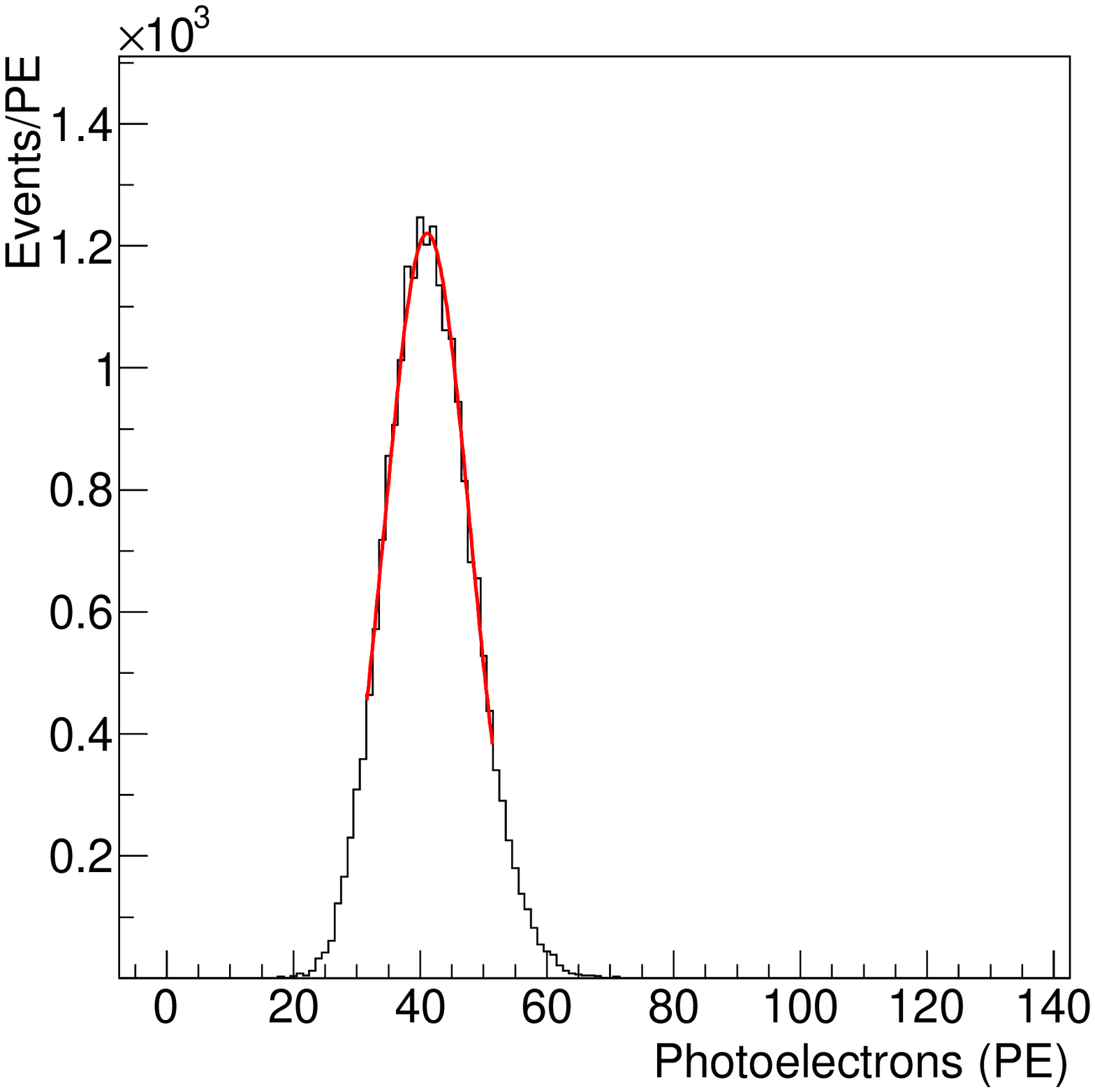}
\includegraphics[scale=0.27]{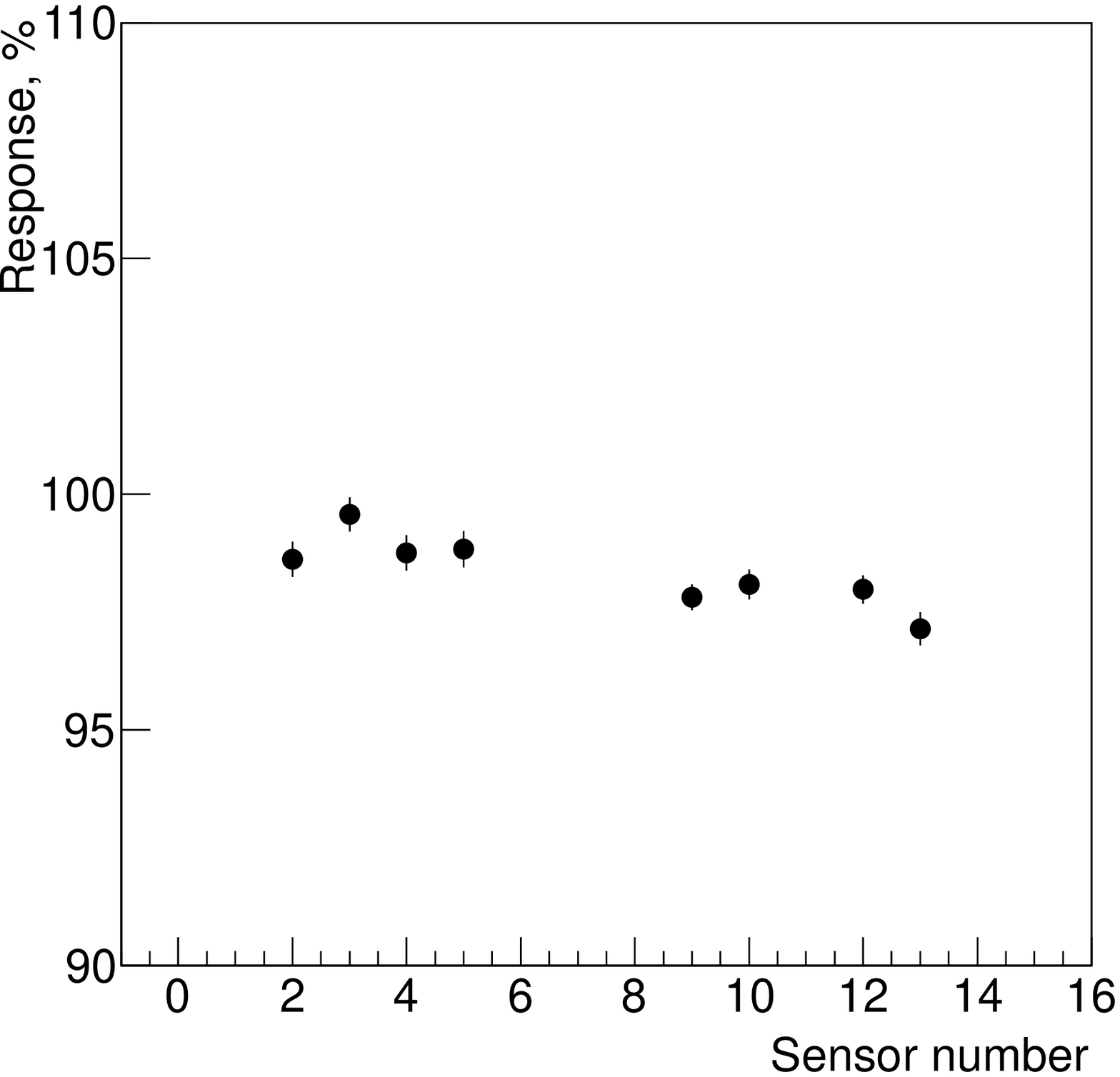}
\includegraphics[scale=0.27]{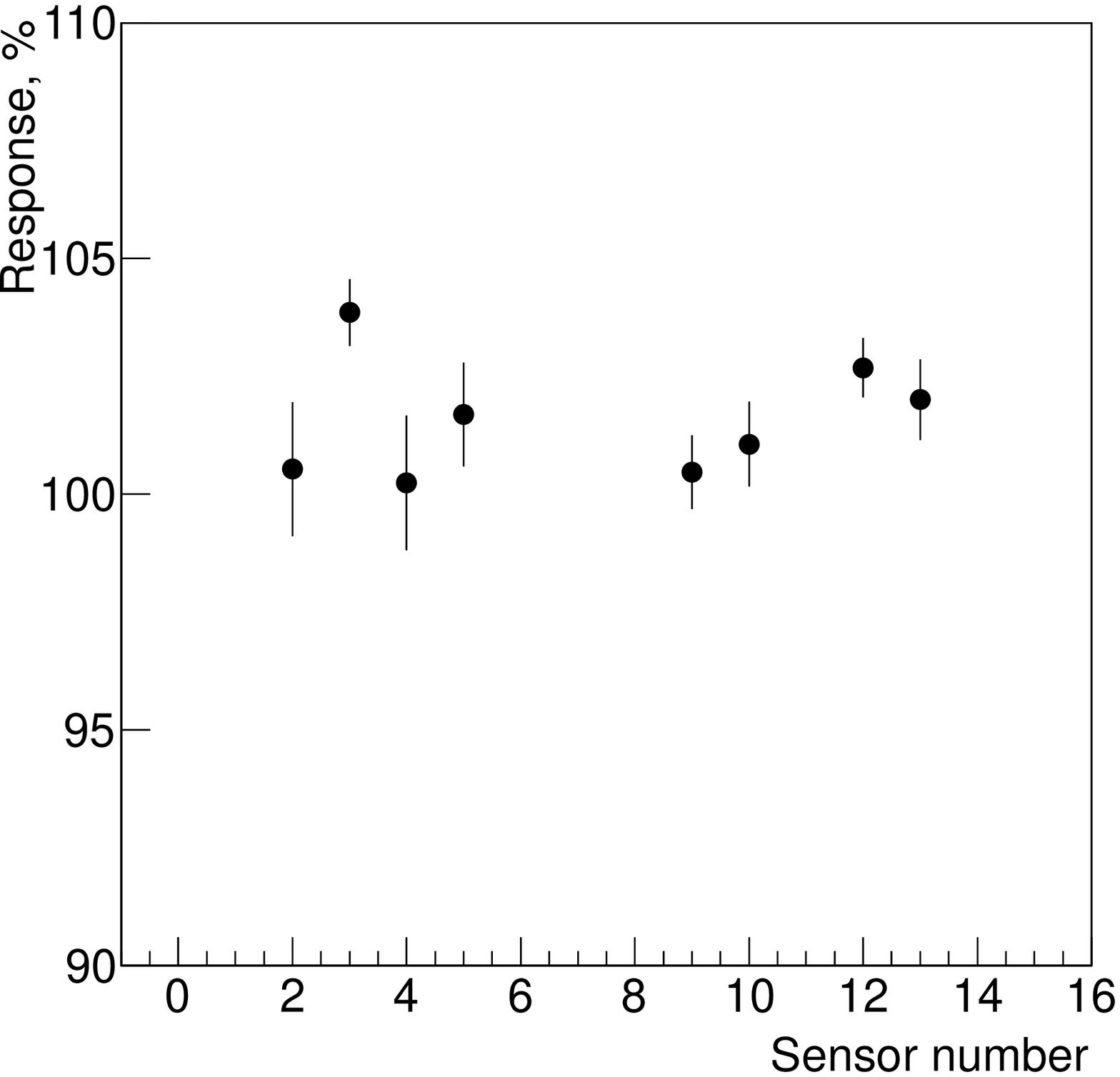}
}
\leftline{ \hspace{2.5cm}{\bf (a)} \hfill\hspace{2.2cm} {\bf (b)} \hfill \hspace{2.5cm} {\bf (c)} \hfill}
\caption{\label{fig:2x2_long_pulse} (a) Non-irradiated MPPC response to a 16~ns  LED signal at V$_{0}$+2.3~V;
(b) irradiated to non-irradiated MPPC response ratio in percent for 5$\times 10^{9}$ p/cm$^2$; 
(c) and for 1$\times 10^{10}$ p/cm$^2$. }
\end{figure}

By exposing the MPPCs to 16 ns LED pulses, we analyzed the effect of radiation on signals with magnitudes of several tens of PE.
Since the 16 ns LED illumination interval is longer than the 12.5 ns FEB sampling period,  we measure the response of the SiPM by summing all five amplitudes in the five-sample region centered on the arrival time of the MPPC signal. The study was done at different irradiation levels.
An example of a pedestal-subtracted, gain-converted response at V$_{0}$+2.3~V is shown in Fig.~\ref{fig:2x2_long_pulse}a. The mean of the Gaussian fit is taken as the response. 

The ratios of the MPPC response at fluences of 5$\times 10^{9}$ p/cm$^2$ and 1$\times 10^{10}$ p/cm$^2$ to the non-irradiated response are shown in Fig.~\ref{fig:2x2_long_pulse}b and  Fig.~\ref{fig:2x2_long_pulse}c, respectively. 
The ratios were put on equal footing by correcting the response and gain for the voltage drop. To within 4\%, the response of the MPPCs remains unchanged after irradiation. The data are not corrected for saturation effects which are on the order of 1\%. 
\section{Summary}
I-V curves, breakdown voltages, dark rates, gains, and LED light response of Hamamatsu 2.0$\times$2.0~mm$^2$ MPPCs were measured before and after irradiation with proton fluences relevant to the Mu2e experiment. No significant deterioration in performance in terms of breakdown voltage, gain, and response is observed. As may be expected, the dark noise rate (and current) increases significantly with radiation, which degrades the ability to identify the single photoelectron peak for fluences exceeding 1$\times 10^{10}$ 1-MeV-equivalent-n/cm$^2$.  

For Mu2e, because of bandwidth limitations the higher noise rate at the highest fluences will require a higher zero-suppression threshold. This raised threshold, however, is still well within the limits required for the veto efficiency.
%
\section{Acknowlegements}
We are grateful for the vital contributions of the Fermilab staff and
the technical staff of the participating institutions. We wish to thank the staff of the Northwestern proton therapy center in Warrenville, IL 
for generous  access to their proton beam. 
This work was supported by the US Department of Energy and the US National Science Foundation.
\end{document}